\documentclass[letterpaper]{article} 
\usepackage{aaai2026}  
\usepackage{times}  
\usepackage{helvet}  
\usepackage{courier}  
\usepackage[hyphens]{url}  
\usepackage{graphicx} 
\usepackage{natbib}  
\usepackage{caption} 
\usepackage{algorithm}
\usepackage{algorithmicx}
\usepackage{algpseudocode}
\usepackage{enumitem}
\usepackage{amsmath}
\usepackage{booktabs}
\usepackage{xcolor}
\usepackage{amssymb}

\usepackage{diagbox}
\usepackage{tcolorbox}
\definecolor{mycolor}{RGB}{20, 0, 145}  
\newtcolorbox{mycolorbox}[1][]{colframe=mycolor, colback=mycolor!4!white, title=#1}

\usepackage{newfloat}
\usepackage{listings}
\DeclareCaptionStyle{ruled}{labelfont=normalfont,labelsep=colon,strut=off} 
\floatstyle{ruled}
\newfloat{listing}{tb}{lst}{}
\floatname{listing}{Listing}
\title{Data Verification is the Future of Quantum Computing Copilots}
\author {
    Junhao Song\textsuperscript{\rm 1}\equalcontrib,
    Ziqian Bi\textsuperscript{\rm 2}\equalcontrib,
    Xinliang Chia\textsuperscript{\rm 3},
    William Knottenbelt\textsuperscript{\rm 1},
    Yudong Cao\textsuperscript{\rm 4}\thanks{Corresponding author.}
}
\affiliations {
    \textsuperscript{\rm 1}Department of Computing, Imperial College London, UK\\
    \textsuperscript{\rm 2}Department of Computer Science, Purdue University, USA\\
    \textsuperscript{\rm 3}Department of Applied Mathematics, National Chung Hsing University, Taiwan\\
    \textsuperscript{\rm 4}BCG X AI Science Institute, Boston Consulting Group, USA\\
    junhao.song23@imperial.ac.uk, bi32@purdue.edu, marcus.chia@ai-agent-lab.com, w.knottenbelt@imperial.ac.uk, cao.yudong@bcg.com
}

\begin{document}

\maketitle

\begin{abstract}
Quantum program generation demands a level of precision that may not be compatible with the statistical reasoning carried out in the inference of large language models (LLMs). Hallucinations are mathematically inevitable and not addressable by scaling, which leads to infeasible solutions. We argue that architectures prioritizing verification are necessary for quantum copilots and AI automation in domains governed by constraints. Our position rests on three key points: verified training data enables models to internalize precise constraints as learned structures rather than statistical approximations; verification must constrain generation rather than filter outputs, as valid designs occupy exponentially shrinking subspaces; and domains where physical laws impose correctness criteria require verification embedded as architectural primitives. Early experiments showed LLMs without data verification could only achieve a maximum accuracy of 79\% in circuit optimization. Our positions are formulated as quantum computing and AI4Research community imperatives, calling for elevating verification from afterthought to architectural foundation in AI4Research.
\end{abstract}


\section{Introduction} \label{sec:intro}
The quantum computing stack requires tools for program \textbf{generation} and program \textbf{compilation} for both near-term physical quantum devices (e.g., superconducting qubits \cite{lu2024camel}, trapped-ion systems \cite{kreppel2024shuttling} and neutral atoms arrays \cite{choi2023preparing,Karuppasamy2025-sx}) and future fault-tolerant quantum computers based on surface codes \cite{google2023suppressing}, qLDPC codes \cite{breuckmann2021quantum} and Bosonic codes \cite{cai2021bosonic}. %
Concurrently, large language models (LLMs) assisted research is becoming increasingly prevalent across various disciplines, but their limitations are gradually emerging within the field of quantum computing. %
Such problems pose formidable challenges for AI automation with a quantum computing copilot: %
\textbf{(1)} They cross multiple layers of abstraction from high-level circuit descriptions to low-level gate arrangements, and \textbf{(2)} have rigorous problem definitions with precise verification methods. %
Critically, they \textbf{(3)} demand zero tolerance for errors, unlike the typical percentage error metrics used for testing LLMs. %

\begin{figure}[t]
    \centering
    \includegraphics[width=\linewidth]{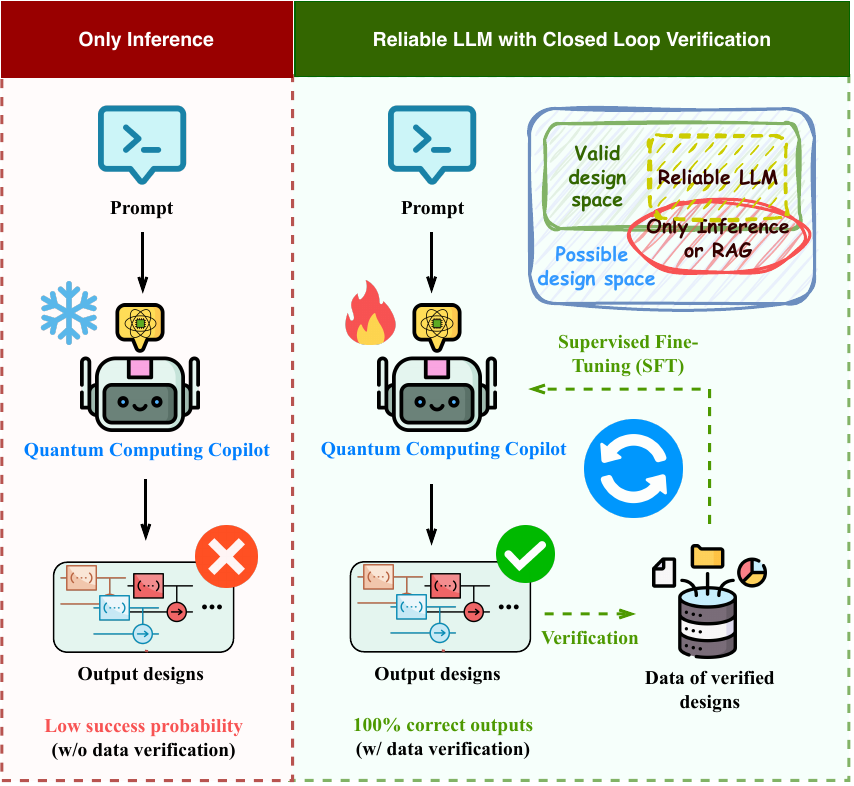}
    \caption{Without data verification, outputs scatter across the exponentially low success probability space (red circle in the top right corner). Even \textbf{Retrieval-Augmented Generation (RAG)} is insufficient to guarantee correctness in this absence. When data verification is integrated into the training loop, only validated designs from the constrained validation space are used, ensuring 100\% correct outputs. The snowflake symbol represents weights are frozen. The flame symbol represents that weights are trainable.}
    \label{fig:agent}
\end{figure}

Reflecting these challenges, synthesizing optimized quantum circuits remains a labor-intensive process that demands manual insights and complex mathematical proofs. %
These processes are inherently inefficient, difficult to formalize, and prone to errors across abstraction layers. %
Consequently, producing an end-to-end optimized design and compilation of a quantum program can often be labor-intensive for a complete algorithm, such as detailed resource estimations for quantum factoring \cite{Gidney2021,Gidney25}. %
The resulting complexity of such human-generated program design and compilation makes these designs challenging for even trained experts to understand and replicate. %
This difficulty motivated the creation of foundational tools including Qualtran \cite{qualtran}, pyLIQTR \cite{pyliqtr}, Bench-Q \cite{BenchQ}, Stim \cite{Gidney2021Stim}, TQEC \cite{TQEC} and Pandora \cite{Pandora}. %
Such tools provide the foundational infrastructure for the development of quantum computing copilots. %
Recent AI systems show promise: IBM's Qiskit Code Assistant outperforms general models on quantum tasks \cite{dupuis2024qiskit}, DeepMind's AlphaTensor-Quantum achieved 37-47\% T-gate reductions \cite{ruiz2025alphatensor}, and Google's AlphaEvolve generated circuit compilation functions for quantum simulation \cite{novikov2025alphaevolve}. These successes reveal LLMs potential to accelerate development cycles and discover non-obvious optimization patterns that human designers might overlook or take longer to produce. %


However, fundamental limitations constrain these approaches. %
LLMs perform pattern matching over training distributions rather than formal reasoning, leading to systematic failures when tasks require multi-step logical operations \cite{dziri2023faith}. %
Theoretical analysis establishes that infeasible solutions are mathematically inevitable in LLMs without data verification, not transient bugs awaiting engineering solutions \cite{karpowicz2025fundamental}. %
For quantum circuits, where correctness requires satisfying rigid mathematical constraints across hundreds or thousands of interdependent gates, statistical improvements in average performance cannot substitute for guaranteed correctness (\textbf{Limitation 1}). %
The insufficient training data problem compounds these quantum copilot issues (\textbf{Limitation 2}). %
The most comprehensive quantum dataset (QDataSet) \cite{perrier2022qdataset} required more than three months (2020-2021) of HPC cluster time to generate 14TB of validated data covering only one-qubit and two-qubits systems. %
Benchmark evaluations reveal that LLMs exhibit consistent error patterns on quantum tasks, with fine-tuning sometimes underperforming few-shot learning \cite{yang2024qcircuitnet}. %
The highly constrained search domain of valid quantum circuits bounded by unitarity, hardware topology and gate fidelity demands verification mechanisms that current LLM architectures cannot provide internally (\textbf{Limitation 3}). %

Considering the three limitations mentioned above, this paper advances three positions: %
\begin{itemize}
    \item \textbf{Data verification as minimum requirement.} Verified datasets ensure copilots internalize constraints as learned structure, not statistical approximation.
    
    \item \textbf{A priori constraints over a posteriori filtering.} Verification must constrain generation; filtering exponentially growing invalid outputs is prohibitive.
    
    \item \textbf{Verification-first paradigm for AI4Research.} Domains with physical laws require verification as architectural primitives in training and generation.
\end{itemize}
The subsequent sections are structured as follows: Section \ref{sec:positions} introduces our positions, Section \ref{sec:arguments} elaborates on both the positions and their supporting arguments, and Section \ref{sec:future} discusses future directions."

\section{Positions} \label{sec:positions}

\textbf{Definition.} We define key terms used throughout this paper:
\begin{itemize}
    \item \textbf{Verification-aware model}: An LLM that has been exposed to formally verified training data where each example satisfies domain-specific correctness constraints (e.g., circuit unitarity, mathematical equivalence proofs). This is distinct from models trained on unverified code repositories.
    \item \textbf{Data verification}: The process of formally proving that training examples satisfy domain constraints using automated theorem provers (Lean, Z3) before inclusion in the training set.
    \item \textbf{A priori constraints}: Verification checks embedded within the generation process that prevent invalid outputs from being produced, as opposed to post-hoc filtering of completed outputs.
\end{itemize}

A central issue concerning the limitation of the current LLMs capabilities is infeasible solutions. 
Regardless of how the size of the model and training data set can be scaled up, infeasible solutions are known to be a fundamental barrier \cite{karpowicz2025fundamental}.
For highly constrained applications such as quantum program generation and compilation, where there is no tolerance for any error, does this mean that LLMs are doomed to fail them?
In this paper, we present our positions that point to a possible way to systemically remove infeasible solutions by creating {\bf verified} datasets. More concretely, we state the positions in this paper as follows:
\begin{enumerate}[label=P\arabic*]
    \item {\bf Data verification must be the minimum requirement for quantum copilots.} This is largely driven by the nature of the problem (Fig. \ref{fig:agent}). Any LLM that has a non-zero chance of producing wrong answers will not be trusted as a viable component of a quantum copilot.
    \item {\bf Verification should provide a priori constraints, not a posteriori filtering.} As the number of design building blocks e.g.\ qubits grows, the fraction of valid designs in the space of all possible designs decreases exponentially (Fig. \ref{fig:agent}). Therefore, any strategy relying on a posteriori filtering will have an exponentially low likelihood of randomly generating a valid design.
    \item {\bf Verification-aware architecture required for highly constrained scientific domains.} We advocate that verified datasets become a common practice in scientific domains that, like quantum computing, can be revolutionized by AI automation but at the same time involve hard constraints such as laws of physics or mathematical axioms. Examples of such scientific disciplines include drug discovery \cite{masters2025investigating}, materials identification \cite{ferguson2025future} and engineering design with physical constraints \cite{wang2025integrating}.
\end{enumerate}

\begin{figure*}
    \centering
    \includegraphics[width=\linewidth]{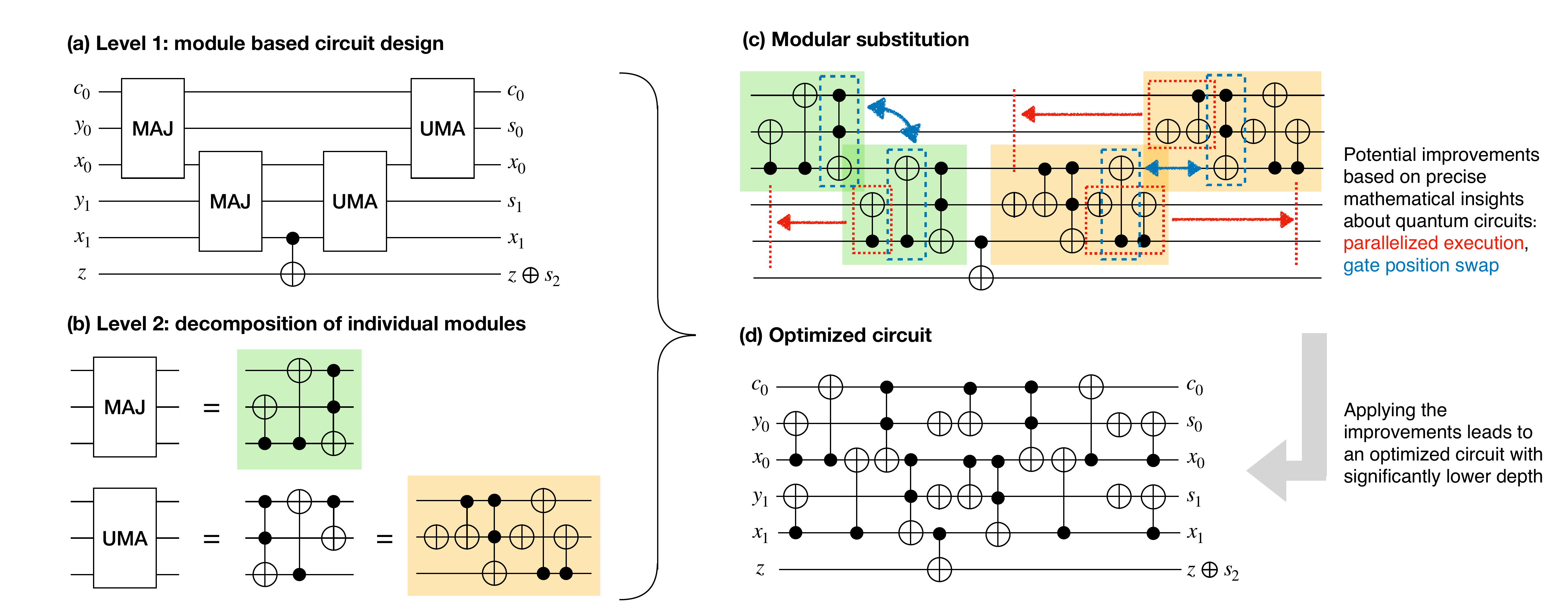}
    \caption{The key features of quantum program \textbf{generation} (step a) and \textbf{compilation} (steps b-d), using the Cuccaro Adder as an example. \textbf{(a)} The first level of abstraction where circuit design is at a modular level. \textbf{(b)} The second level where each module is further decomposed into elementary gates (NOT, CNOT and Toffoli). \textbf{(c)} Expanding the modules to reveal the overall circuit design with only the elementary gates, and identifying opportunities for improvements. \textbf{(d)} Applying the improvements (parallelized execution of gates and gate position swap due to commutativity) yields an optimized circuit with lower depth. }
    \label{fig:cuccaro_adder}
\end{figure*}

\section{Position Arguments} \label{sec:arguments}
We first describe the formulation of the program generation and compilation problem, using Cuccaro Adder \cite{Cuccaro2004ripple} as an example. The adder is an in-place quantum program that performs the addition of two $n$-bit integers $x$ and $y$. The transformation of the adder can be described as:
\begin{equation}\label{eq:adder_req}
    |c_0\rangle|x\rangle|y\rangle|0\rangle\mapsto|c_0\rangle|x\rangle|x+y\rangle,
\end{equation}
where $c_0$ is an incoming carry bit and the extra ancilla in $|0\rangle$ captures the overflow in the addition process. Any adder design must exactly satisfy the requirement described in Equation \ref{eq:adder_req} or any equivalent statement. This places a constraint of \textbf{precise mathematical specification} that may not apply to other AI automation tasks. 

In Fig. \ref{fig:cuccaro_adder}a we show an example for $n=2$, where the adder is constructed by a cascade of Majority (MAJ) modules followed by a cascade of UnMajority-and-Add (UMA) modules. This is the program generation step. The generated problem can be further compiled by expanding the modules with elementary gates (Fig. \ref{fig:cuccaro_adder}b), substituted into the original circuit for further improvement (Fig. \ref{fig:cuccaro_adder}c) yield an optimized circuit (Fig. \ref{fig:cuccaro_adder}d) that breaks the modular design in the original iteration. Such a pattern of \textbf{leaky abstraction} is common in quantum program generation and compilation, where two levels of description (such as Level 1 in Fig. \ref{fig:cuccaro_adder}a and Level 2 in Fig. \ref{fig:cuccaro_adder}b) need to be merged (``leaked'') into each other to produce an optimized implementation. A consequence of such leaky abstraction is {\bf non-decomposibility} of the design space, namely that suboptimal designs at a certain level of abstraction (such as the highlighted design on UMA in Fig.\ \ref{fig:cuccaro_adder}b, which is arguably less optimal than the other alternative with only 3 gates) do not necessarily yield suboptimal designs when expanding into lower layers of abstraction. The highlighted design of UMA in Fig. \ref{fig:cuccaro_adder}b turns out to be more advantageous in the overall circuit optimization by parallelizing gate executions \cite{Cuccaro2004ripple}. The improvements shown in Fig. \ref{fig:cuccaro_adder} are but a subset of improvements that can be made \cite{Cuccaro2004ripple}. However, finding those detailed improvements with leaky abstraction is largely a task carried out by human experts. In 2018, Craig Gidney \cite{gidney2018halving} is another one such example in a different context (fault-tolerant quantum implementation), but also for the Cuccaro adder. As such, the task of quantum program generation and compilation is still a \textbf{labor-intensive} effort that highly depends on the artisanal skills of human expert.

Faced with the aforementioned key features of the program generation and compilation problem, namely precision in the mathematical specification, leaky abstraction of the design space, and labor intensity of the design task, it is important to be intentional in the design of any AI automation systems. To that end, we expand on the positions laid out in Section \ref{sec:positions} as the following.

\begin{figure}[htbp]
    \centering
    \includegraphics[width=\linewidth]{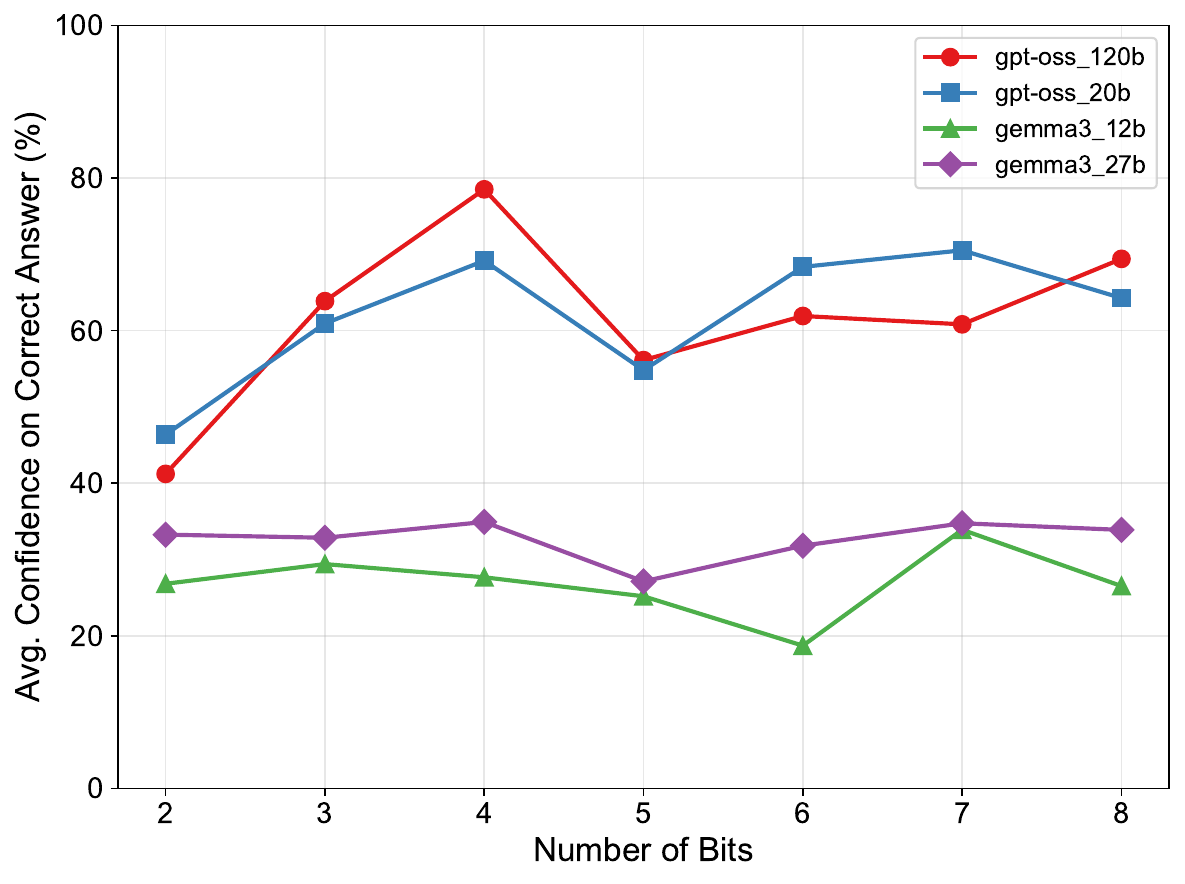}
    \caption{Average confidence assigned to the correct answer for the top-5 performing models. Higher values indicate better calibration. GPT-OSS models show higher confidence on correct answers compared to Gemma3 models, suggesting better uncertainty quantification. Note: \texttt{qwen3:1.7b}, despite ranking among the top-5 in accuracy, is omitted because it fails to produce valid token-level probabilities (outputting malformed responses that yield zero confidence scores).}
    \label{fig:calibration}
\end{figure}

\textbf{(P1) Data verification must be the minimum requirement for quantum copilots, due to the limitation of current LLMs in understanding the structure of quantum circuit compilations.} The design of quantum circuits comes with exact constraints that need to be satisfied. In the case of arithmetic circuit design such as the quantum adder, unitarity of the circuit operation implies that the logical operation implemented by the circuit needs to be reversible. How to ensure that the AI system internalizes such precise constraints? Our position is that verified training data ensures that the model observes only reversible, resource-annotated circuits; without it the model cannot learn the precise mathematical constraints that underpins the quantum program generation and compilation task.

Training quantum copilots on unverified data from sources such as low quality or uncommented code constitutes implicit data poisoning. Unlike natural language tasks where massive data volumes average out noise to approximate ground truth, quantum circuit constraints do not admit this: a single invalid gate sequence violates unitarity regardless of how many valid examples surround it in the solution space. This creates an anti-scaling phenomenon where additional unverified data amplifies rather than dilutes errors.

To substantiate our position we derived an evaluation suite directly from the verified MAJ/UMA corpus described earlier.

\textbf{Experimental Setup.} We evaluated 34 publicly available LLMs spanning 270M to 120B parameters, deployed via Ollama with default parameters. Each model answered 70,000 multiple-choice questions (10,000 per bit-width from 2 to 8 qubits). The evaluation harness includes a retry mechanism (up to 3 attempts) for malformed responses, with unanswered questions counted as incorrect.

\begin{figure}[t]
    \centering
    \includegraphics[width=\linewidth]{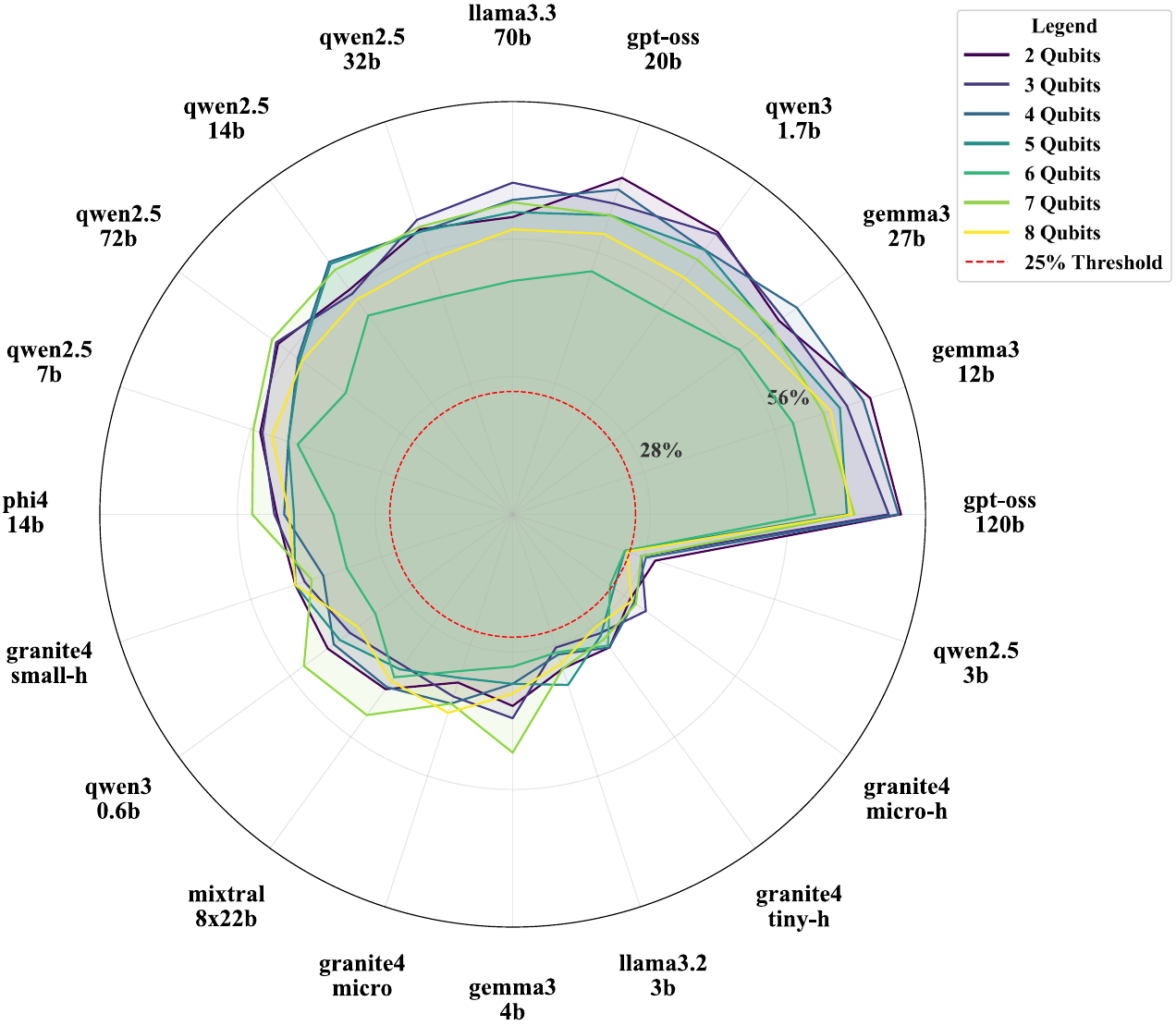}
    \caption{\textbf{Average accuracy per LLMs}. The dashed circle marks the 25\% random baseline; verification-aware models clearly exceed it, while unconstrained checkpoints remain inside.}
    \label{fig:radar}
\end{figure}

For every $n \in \{2,\dots,8\}$ we construct 10,000 multiple-choice questions where four candidate circuits all satisfy the adder specification but differ in gate count, depth, and Toffoli usage. The ground-truth answer is the circuit with the minimum score which is a weighted combination of circuit parameters
$
    0.5 \times \mathrm{Toffoli}_{n} + 0.25 \times \mathrm{Depth}_{n} + 0.25 \times \mathrm{TotalGates}_{n},
$
so that each decision compares equally valid but differently optimized implementations. Models with verification knowledge (\texttt{gemma3:12b}, \texttt{gpt-oss:120b}) achieve 0.60-0.79 accuracy, while generic models hover near a random baseline (0.21-0.29). The radar view in Figure~\ref{fig:radar} compresses these results by averaging each model across bit-widths: these \textbf{verification-aware models} dominate the outer ring, while unconstrained copilots cluster inside the dashed 25\% circle. This quantitative gap demonstrates that scaling parameter counts alone cannot guarantee correctness without verified data and structural constraints.

Beyond accuracy, we examine model calibration (Figure~\ref{fig:calibration}). For each question, we extract the log-probabilities of tokens corresponding to the four options (A, B, C, D) from the model's output distribution. Since raw log-probabilities do not sum to a consistent value across the four options, we apply softmax normalization to obtain calibrated confidence scores that form a proper probability distribution. GPT-OSS models consistently assign 60--80\% confidence to correct responses across all bit-widths, while Gemma3 models remain in the 20--35\% range. This disparity suggests that verification-aware training not only improves accuracy but also enhances uncertainty quantification: models exposed to verified data exhibit more appropriate confidence when producing correct answers.

Several models (e.g., \texttt{mixtral:8x7b}, \texttt{deepseek-r1} variants) scored below the 25\% random baseline due to malformed outputs—verbose explanations or refusals rather than single-letter answers. This failure mode reinforces our position: models without verification-aware training cannot reliably engage with formally constrained tasks.


\textbf{(P2) Verification should provide a priori constraints, not a posteriori filtering, due to domain constraints in quantum circuit search spaces.} Models that internalize explicit penalties for depth and Toffoli count rarely emit disallowed options, whereas unconstrained models require aggressive post-hoc rejection sampling that wastes compute and still leaks invalid answers. Consider the adder design space: with $d$ gate choices per position and $O(n)$ positions, the total space grows as $O(d^n)$. Constraints reduce valid designs to $O(\delta^n)$ for $\delta<d$, yielding a valid fraction of $O((\delta/d)^n)$ that decreases exponentially. Post-hoc filtering thus incurs exponential cost.

Concretely, to demonstrate our position, we have generated a dataset of over 200,000 adder circuit designs all of which implement the transformation in Equation \ref{eq:adder_req} for any sized input $x$ and $y$. Each design in the dataset is formally verified with Lean to show correctness for any $n$ and with Z3 to show correctness for modular level decomposition of MAJ and UMA operations. Through a process that combines exhaustive search and formal verification (with Lean and Z3), we discovered 540 MAJ circuit designs and 529 UMA circuit designs by searching over circuit depths 3 through 6. The combination of the designs leads to over $540\times 529=285660$ distinct adder circuit designs. However, this number is still a negligible fraction of possible circuit designs. Each MAJ/UMA module is in a space of approximately 12.2 million gate arrangements across depths 3-6. The combined design space exceeds 148 trillion configurations, yet formal verification identifies only over 200,000 valid adders (a ratio on the order of $10^{-9}$). This exponential gap between possible and valid designs renders post-hoc filtering computationally prohibitive.


Recent systems adopt this principle: QCircuitNet pairs tasks with executable oracles \cite{yang2024qcircuitnet}, AlphaTensor builds correctness into search objectives \cite{fawzi2022discovering}, and AlphaEvolve filters variants through machine-checkable specifications \cite{novikov2025alphaevolve}, demonstrating that embedding verification within generation loops enables scalable discovery.

\textbf{(P3) Verification-aware architecture is required not only for quantum circuit design but also for other problems throughout science.} The design of the adder circuit in Fig. \ref{fig:cuccaro_adder} is at the level of logical circuit design. It can be lowered into a more detailed compilation towards a specific architecture of fault tolerant quantum hardware, such as 2D surface code \cite{gidney2018halving}. In this case, a similar leaky abstraction and non-decomposability apply between the logical quantum circuit and its physical implementation on a fault tolerant quantum computer. 

A general formulation of such leaky abstraction is a setting where one chooses implementations for components $g_1,\dots,g_m$ and then compose them into a system scored by a global objective $f$. Non-decomposability then manifests even if each $g_k$ admits a well-defined notion of ``locally optimal'' design, the composed objective $f$ is not generally equal to a sum of independent per-module objectives over disjoint variable blocks, so independently maximizing (or minimizing) each $g_k$ does not guarantee optimality for $f$. 

The same non-decomposability appears broadly: in drug discovery, optimizing a single property (e.g., potency) can worsen others (e.g., permeability, metabolic stability, safety), motivating explicit {multi-parameter optimization} scores that trade off multiple physicochemical attributes rather than treating them independently \cite{Wager2010CNSMPO}. In combinatorial optimization, greedy heuristics that make the best local choices can perform arbitrarily poorly on global objectives (e.g., for Traveling Salesperson Problem (TSP) variants there are instances where greedy-type tour construction returns the worst tour), illustrating that ``locally favorable'' steps need not make up a globally good solution \cite{GutinYeoZverovich2001}. In engineering design, multidisciplinary systems couple aerodynamics, structures, controls, thermal behavior, etc., so improving one discipline in isolation can degrade overall system performance or feasibility; this motivates multidisciplinary design optimization architectures that coordinate interacting subsystems under a global objective and shared constraints \cite{MartinsLambe2013MDO}.


At the ecosystem level, robustness and generalization hinge on community agreed benchmarks, standardized evaluation metrics, curated archives with stable metadata/interfaces and uncertainty quantification, such as verification practices elevated from individual labs to shared infrastructure, so that models are developed and assessed against domain-meaningful, reproducible criteria rather than ad-hoc proxies \cite{ferguson2025future}.


\section{Future Directions} \label{sec:future}
Two main research directions follow from our verification first paradigm:
\begin{itemize}
    \item \textbf{FD1} Quantum copilots architectures must integrate formal verification directly into the LLMs token generation loop rather than treating it as post-hoc validation (details below). 
    \item \textbf{FD2} The community needs constraint rich benchmarks, releasing such datasets under open licenses would establish reproducible standards for quantum copilot development and enable rigorous cross system comparison.
\end{itemize}

For the \textbf{FD1}, we suggest that each proposed circuit fragment should satisfy \textbf{satisfiability modulo theories constraint} or symbolic specifications before being committed to the design. So that can transform verification from a binary filter into a structured exploration guide that avoids invalid paths in quantum copilot's real-time generation. 

For the \textbf{FD2}, the constraint rich benchmarks that go beyond the quantum adder (this paper example) adder evaluation to encompass modular multiplication, quantum Fourier transforms, hardware specific routing constraints and zero tolerance resource costs. Also, we suggest each benchmark can paired to \textbf{executable verification} oracles that assess both functional correctness and optimization quality.

While our arguments focus on quantum circuit optimization, the underlying principles apply broadly: any domain where (1) correctness is binary rather than approximate, (2) valid solutions occupy an exponentially small subspace, and (3) formal verification methods exist, will benefit from verification-first AI architectures. Quantum computing serves as an ideal testbed because mature verification tools and well-defined correctness criteria (unitarity, gate equivalence) already exist. 

More broadly, the data verification principles we demonstrate: embedding domain constraints within objectives and generation loops. Apply across scientific disciplines governed by physical laws or mathematical axioms (not only the quantum area), from drug discovery's biochemical plausibility requirements to materials design's thermodynamic stability constraints. Those different areas can build data verification with constraint specification and solution checking would helps AI better understand the laws of the physical/real world, leveraging the formal methods each domain has developed over decades as foundations for reliable AI-assisted discovery.

\bibliography{aaai2026}



\twocolumn[{
\renewcommand\twocolumn[1][]{#1}
\section*{Data Verification is the Future of Quantum Computing Copilots \\ Supplementary Material}
\vspace{5mm}
}]
\appendix


\section*{Prompt Template} \label{app:prompt}

All experiments reuse the same prompt, implemented verbatim inside the evaluation harness. Angle-bracket placeholders are filled with the MAJ/UMA sequences for each option before the request is sent to LLMs.

\begin{mycolorbox}[Multiple-choice Problem Prompt]
Select the best quantum circuit implementation with the
LOWEST computational cost for an $n$-bit adder.

\textbf{Background:}
This circuit implements an $n$-bit Quantum Ripple-Carry Adder. The architecture is constructed by cascading multiple instances of these two primary modules:
\begin{itemize}
    \item \textbf{MAJ (Majority):} A gate sequence used to propagate the carry bit. In an $n$-bit adder, MAJ blocks are cascaded from the least significant bit to the most significant bit.
    \item \textbf{UMA (UnMajority and Add):} The inverse operation of MAJ, combined with an addition step. These are cascaded in reverse order to compute the sum and uncompute the carry bits.
\end{itemize}

\textbf{Evaluation Criteria:}
\begin{itemize}
    \item $\text{SCORE} = 0.5 \times \text{TOFFOLI\_COUNT} + 0.25 \times \text{CIRCUIT\_DEPTH} + 0.25 \times \text{TOTAL\_GATES}$
    \item Lower SCORE means better performance.
\end{itemize}

\textbf{Options:}
\newline
\textit{Option A:}
\newline
\textless gate\_sequence\_A \textgreater
\newline
\textit{Option B:}
\newline
\textless gate\_sequence\_B \textgreater
\newline
\textit{Option C:}
\newline
\textless gate\_sequence\_C \textgreater
\newline
\textit{Option D:}
\newline
\textless gate\_sequence\_D \textgreater
\newline

IMPORTANT: Output ONLY the letter (A, B, C, or D).
Your answer must be exactly one letter without punctuation.

If uncertain, select the option that appears most optimal.

Your answer:
\end{mycolorbox}

\section*{Example Evaluation Instance} \label{app:casestudy}

The following presents a representative evaluation instance for inputs of size $n=2$ bits. The correct answer is determined by computing the weighted score for each option.

\vspace{0.5em}
\noindent\textbf{Question Instance (2-bit Adder)}

\vspace{0.3em}
\noindent\fbox{\parbox{0.97\columnwidth}{
\small
\textbf{Option A} \hfill \colorbox{green!20}{\textbf{$\checkmark$ Ground Truth}}  \\[2pt]
\texttt{CNOT(1,0) $\circ$ CNOT(2,1) $\circ$ CNOT(1,2) $\circ$ Toffoli(1,0,2) $\circ$ CNOT(1,0) $\circ$ CNOT(3,2) $\circ$ CNOT(4,3) $\circ$ CNOT(3,4) $\circ$ Toffoli(3,2,4) $\circ$ CNOT(3,2) $\circ$ CNOT(4,5) $\circ$ CNOT(3,4) $\circ$ Toffoli(3,2,4) $\circ$ CNOT(3,4) $\circ$ CNOT(4,2) $\circ$ CNOT(2,3) $\circ$ CNOT(1,2) $\circ$ Toffoli(1,0,2) $\circ$ CNOT(1,2) $\circ$ CNOT(2,0) $\circ$ CNOT(0,1)}

\vspace{4pt}\hrule\vspace{4pt}

\textbf{Option B} \\[2pt]
\texttt{X(0) $\circ$ CNOT(2,0) $\circ$ CNOT(2,1) $\circ$ CNOT(1,2) $\circ$ Toffoli(1,0,2) $\circ$ X(0) $\circ$ X(2) $\circ$ CNOT(4,2) $\circ$ CNOT(4,3) $\circ$ CNOT(3,4) $\circ$ Toffoli(3,2,4) $\circ$ X(2) $\circ$ CNOT(4,5) $\circ$ Toffoli(3,2,4) $\circ$ CNOT(2,3) $\circ$ CNOT(4,2) $\circ$ SWAP(4,3) $\circ$ CNOT(4,3) $\circ$ CNOT(3,4) $\circ$ Toffoli(1,0,2) $\circ$ CNOT(0,1) $\circ$ CNOT(2,0) $\circ$ SWAP(2,1) $\circ$ CNOT(2,1) $\circ$ CNOT(1,2)}

\vspace{4pt}\hrule\vspace{4pt}

\textbf{Option C} \\[2pt]
\texttt{CNOT(1,2) $\circ$ CNOT(2,1) $\circ$ SWAP(2,1) $\circ$ CNOT(2,0) $\circ$ Toffoli(0,1,2) $\circ$ CNOT(3,4) $\circ$ CNOT(4,3) $\circ$ SWAP(4,3) $\circ$ CNOT(4,2) $\circ$ Toffoli(2,3,4) $\circ$ CNOT(4,5) $\circ$ CNOT(3,2) $\circ$ CNOT(2,4) $\circ$ Toffoli(3,2,4) $\circ$ CNOT(4,3) $\circ$ CNOT(3,2) $\circ$ SWAP(4,2) $\circ$ CNOT(1,0) $\circ$ CNOT(0,2) $\circ$ Toffoli(1,0,2) $\circ$ CNOT(2,1) $\circ$ CNOT(1,0) $\circ$ SWAP(2,0)}

\vspace{4pt}\hrule\vspace{4pt}

\textbf{Option D} \\[2pt]
\texttt{CNOT(2,0) $\circ$ SWAP(2,1) $\circ$ CNOT(2,1) $\circ$ Toffoli(0,1,2) $\circ$ CNOT(1,2) $\circ$ CNOT(4,2) $\circ$ SWAP(4,3) $\circ$ CNOT(4,3) $\circ$ Toffoli(2,3,4) $\circ$ CNOT(3,4) $\circ$ CNOT(4,5) $\circ$ CNOT(2,3) $\circ$ X(3) $\circ$ Toffoli(3,2,4) $\circ$ X(3) $\circ$ CNOT(4,3) $\circ$ CNOT(4,2) $\circ$ CNOT(0,1) $\circ$ X(1) $\circ$ Toffoli(1,0,2) $\circ$ X(1) $\circ$ CNOT(2,1) $\circ$ CNOT(2,0)}
}}

\vspace{0.8em}
\noindent\textbf{Observed Failure Modes}
\vspace{0.2em}

\begin{itemize}[leftmargin=1.2em, itemsep=2pt, topsep=0pt]
    \item \textit{Format failure:} Models output verbose explanations instead of a single letter.
    \item \textit{Reasoning error:} Models select valid but suboptimal circuits.
    \item \textit{Refusal:} Models decline to answer domain-specific questions.
\end{itemize}

\section*{Details and Visualizations} \label{app:pseudocode&visual}

Algorithm \ref{alg:evaluation} outlines the robust evaluation harness, including the input sanitization and retry logic. This pseudocode details the core logic used to test each LLMs for our early experiments, including the $\textit{MAX\_RETRIES}$ loop to handle generation failures and the \Call{Sanitize}{} procedure used to normalize and validate the raw LLM outputs.

\textbf{Reproducibility Details.}
\begin{itemize}
    \item \textbf{Models}: All 34 models are publicly available through Ollama (https://ollama.ai).
    \item \textbf{Inference}: Default Ollama parameters with max\_tokens sufficient for single-letter responses.
    \item \textbf{Dataset}: 70,000 multiple-choice questions generated from 2M+ formally verified circuit designs.
    \item \textbf{Verification}: Circuits verified using Lean 4 for functional correctness and Z3 for modular decomposition.
    \item \textbf{Evaluation}: Each question has exactly one correct answer determined by the scoring function.
    \item \textbf{Hardware}: Experiments conducted on Apple M3 with 512GB unified memory.
\end{itemize}

This section also presents supplementary visualizations to illustrate the results of our early experiments further, showing LLMs' performance to support our positions in this position paper.

\begin{algorithm}[ht]
\caption{Evaluation Agent Pseudocode}
\label{alg:evaluation}
\begin{algorithmic}[1]

\Procedure{EvaluateHarness}{$Q, \text{model}$}
    \State $\textit{MAX\_RETRIES} \gets 3$
    \State $results \gets \{\}$ \Comment{Initialize results dictionary}
    
    \For{each question $q \in Q$}
        \State $prompt \gets \Call{RenderPrompt}{q}$ \Comment{Render prompt with MAJ/UMA sequences}
        \State $answered \gets \textbf{false}$
        
        \For{$\textit{attempt} \gets 1$ \textbf{to} $\textit{MAX\_RETRIES}$}
            \State $\textit{raw} \gets \Call{CallVLLM}{\text{model}, prompt}$
            \State $\textit{answer} \gets \Call{Sanitize}{\textit{raw}}$
            
            \If{$\textit{answer} \in \{\text{'A'}, \text{'B'}, \text{'C'}, \text{'D'}\}$}
                \State $\Call{RecordCorrectness}{results, q, \textit{answer}}$
                \State $answered \gets \textbf{true}$
                \State \textbf{break} \Comment{Exit retry loop}
            \Else
                \State $\Call{LogFormatError}{q, \textit{attempt}, \textit{raw}}$
            \EndIf
        \EndFor
        
        \If{\textbf{not} $answered$}
            \State $\Call{MarkAsUnanswered}{results, q}$ \Comment{Counts as incorrect}
        \EndIf
    \EndFor
    
    \State \textbf{return} $results$
\EndProcedure

\Statex \Comment{Helper function for sanitizing vLLM output}
\Procedure{Sanitize}{$\textit{raw}$}
    \State $\textit{clean} \gets \Call{TrimWhitespace}{\textit{raw}}$
    \State $\textit{clean} \gets \Call{StripPunctuation}{\textit{clean}}$ \Comment{e.g., quotes, brackets}
    \State $\textit{clean} \gets \Call{Uppercase}{\textit{clean}}$
    \State \textbf{return} $\textit{clean}$ \Comment{Return first token if needed}
\EndProcedure

\end{algorithmic}
\end{algorithm}

\begin{figure*}[t]
    \centering
    \includegraphics[width=\linewidth]{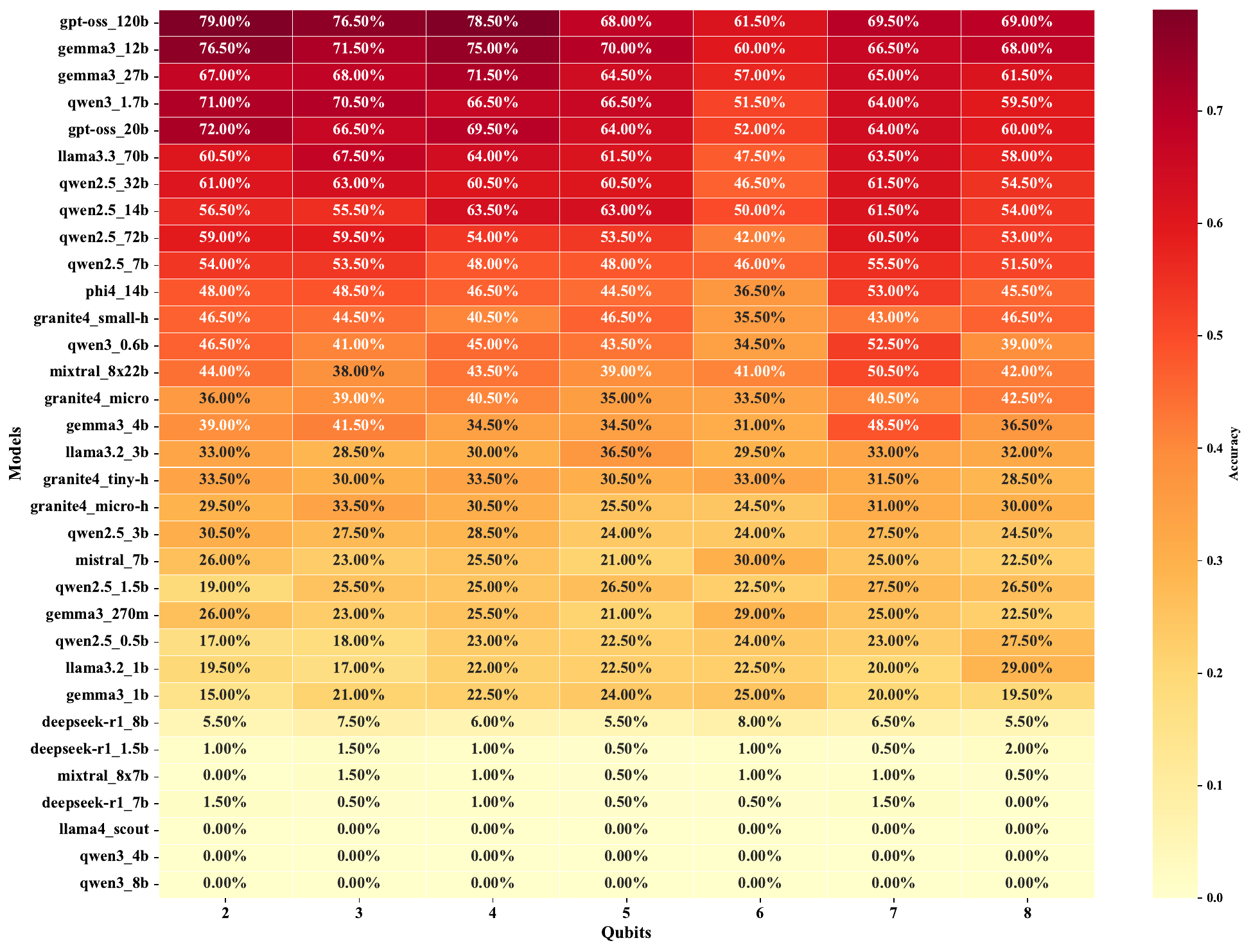}
    \caption{Accuracy heatmap across 34 models and seven bit-widths. Dark cells highlight specific (quantum) verification knowledge LLMs, while lighter bands reveal general-purpose models stuck near random (25\%).}
    \label{app:heatmap}
\end{figure*}

\begin{figure*}[t]
    \centering
    \includegraphics[width=0.48\textwidth]{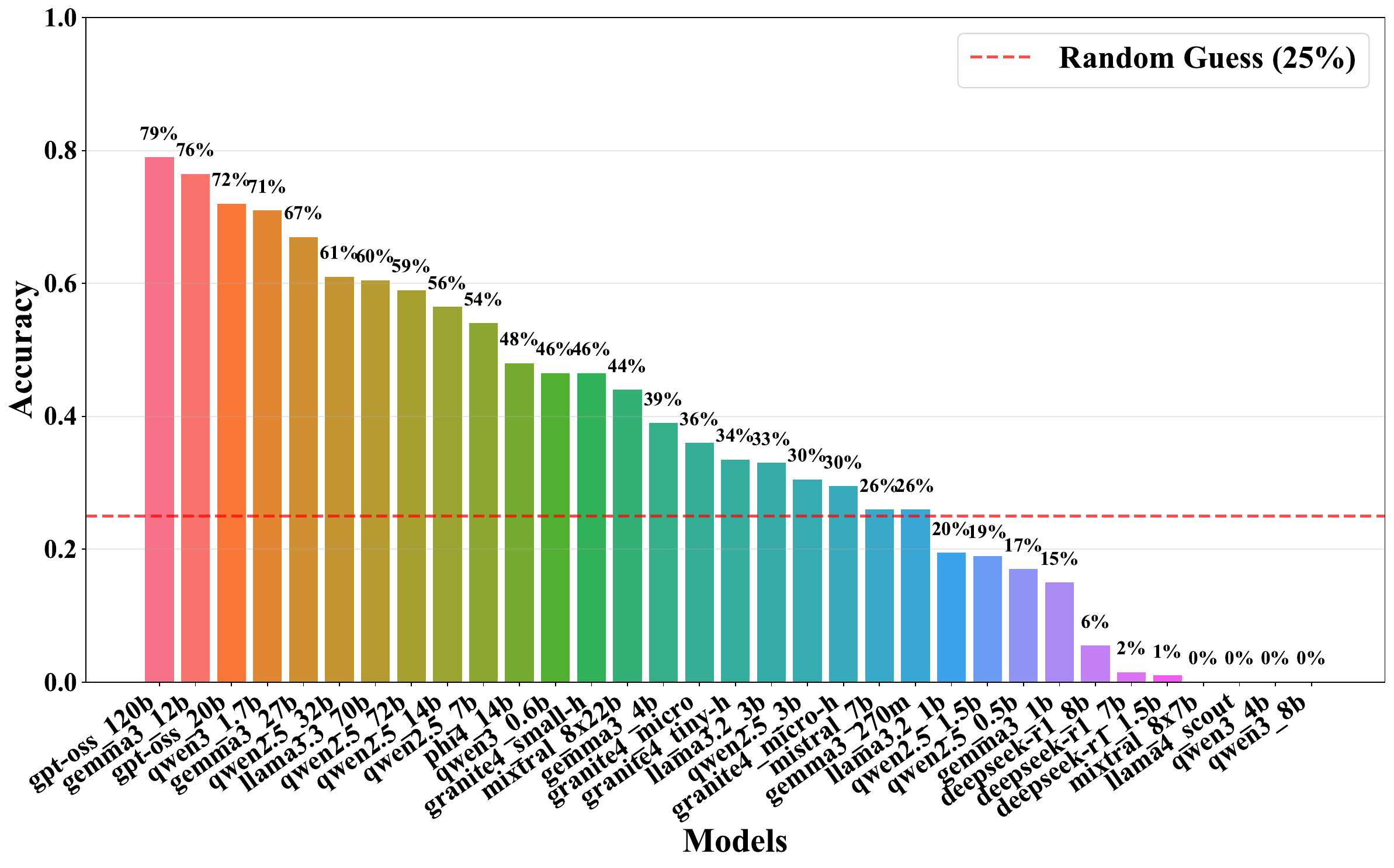}
    \hfill
    \includegraphics[width=0.48\textwidth]{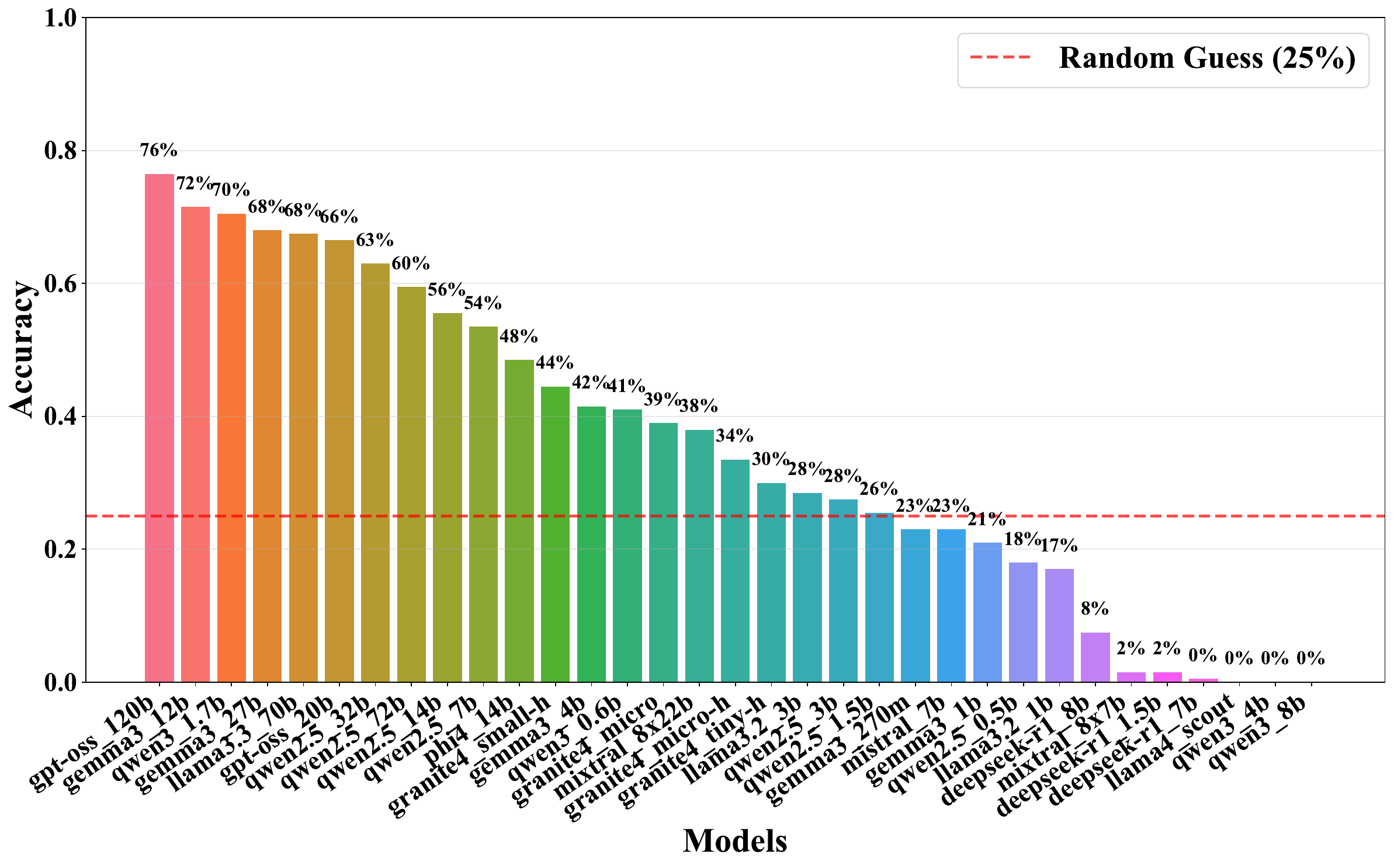}
    \caption{Accuracy rankings for the 2- and 3-qubit MAJ/UMA prompts; both panels sort all 34 copilots and highlight the 25\% random-guess baseline, showing that verification-conditioned checkpoints (e.g., \texttt{gpt-oss:120b}, \texttt{gemma3:12b/27b}, \texttt{qwen2.5:14b} and larger) already sustain 0.6--0.8 accuracy while conventional Gemma, Llama, DeepSeek, and Mixtral variants remain clustered around chance.}
    \label{fig:bit-row1}
\end{figure*}

\begin{figure*}[t]
    \centering
    \includegraphics[width=0.48\textwidth]{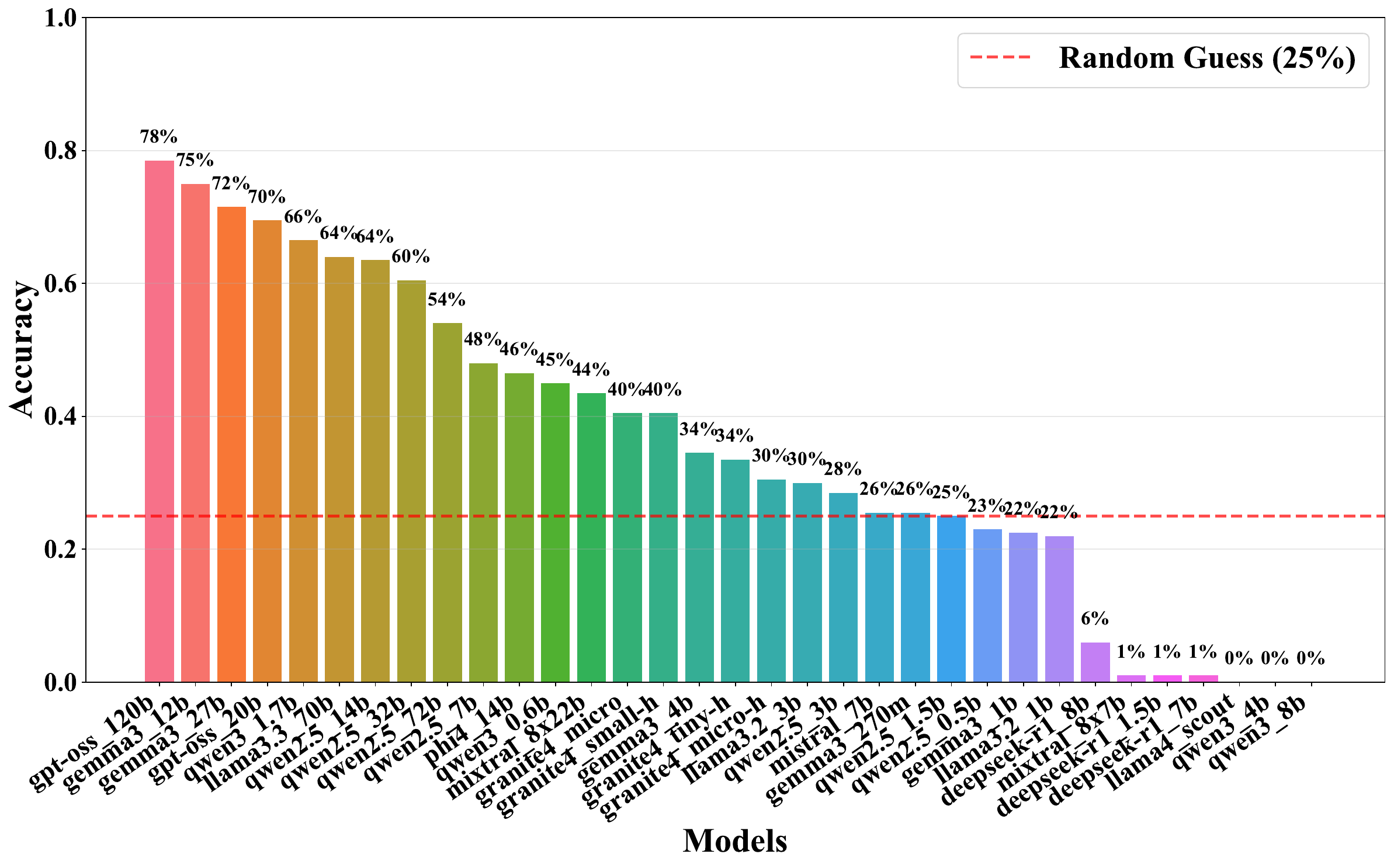}
    \hfill
    \includegraphics[width=0.48\textwidth]{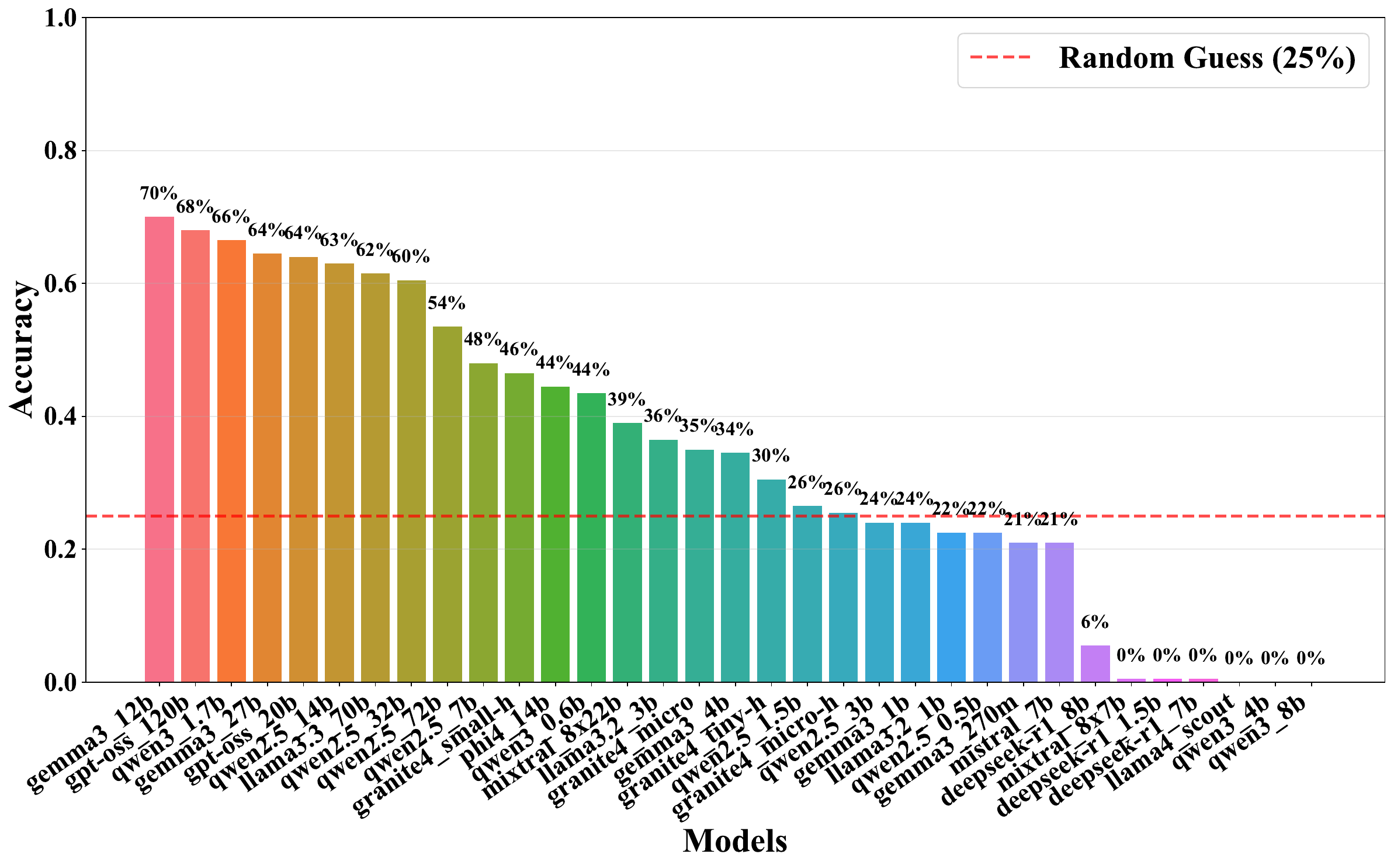}
    \caption{Accuracy rankings for the 4- and 5-qubit settings; verification-aware checkpoints stay above 0.6 despite deeper circuits, whereas large yet unverified copilots plateau near 0.5 on the shallower cases.}
    \label{fig:bit-row2}
\end{figure*}

\begin{figure*}[t]
    \centering
    \includegraphics[width=0.48\textwidth]{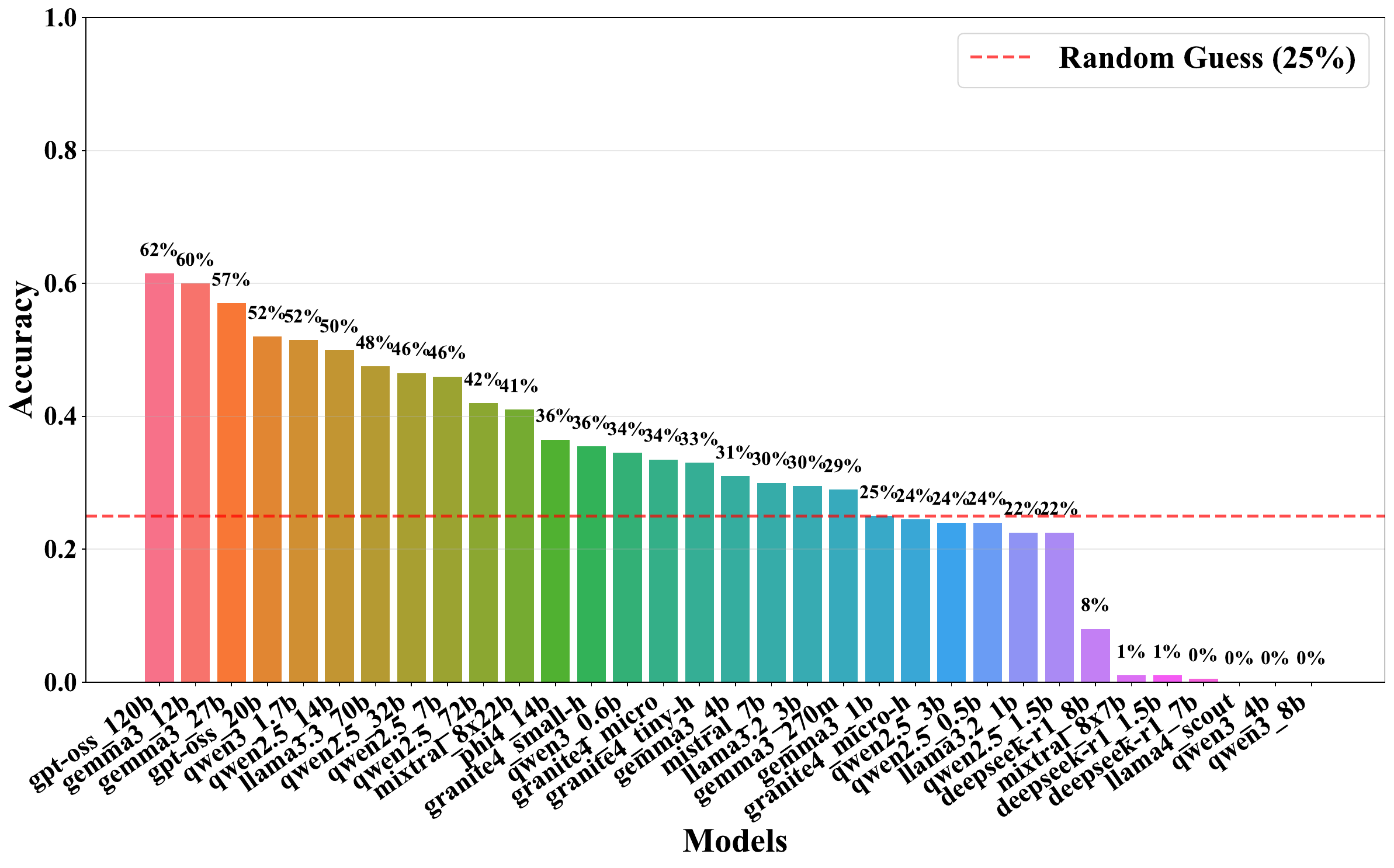}
    \hfill
    \includegraphics[width=0.48\textwidth]{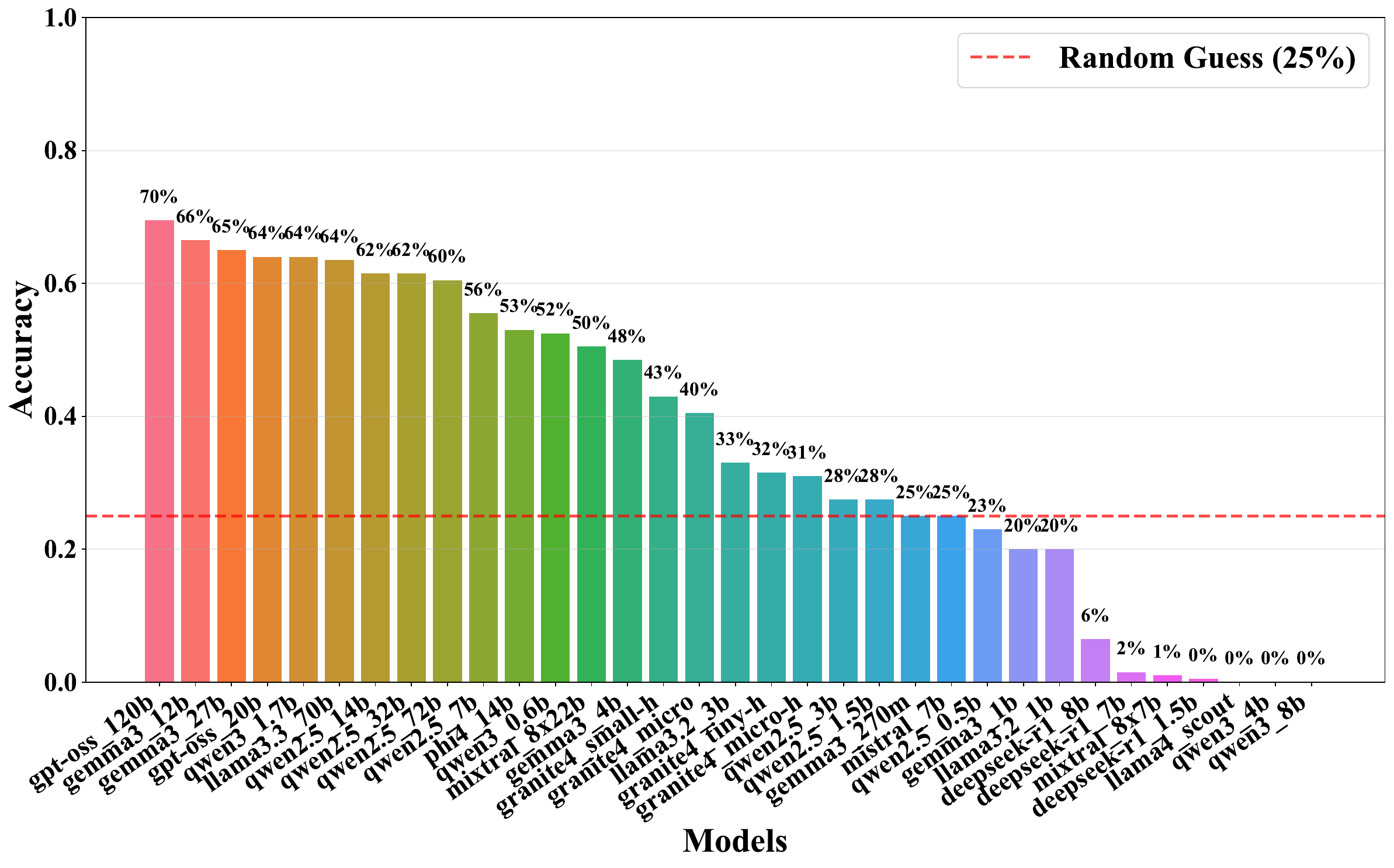}
    \caption{Accuracy rankings for the 6- and 7-qubit benchmarks; verification-aware checkpoints remain above 50\% while unverified copilots (e.g., \texttt{llama3.3:70b}, \texttt{mixtral:8x22b}) slide into the mid-30\% range with most smaller models collapsing outright.}
    \label{fig:bit-row3}
\end{figure*}

\begin{figure*}[t]
    \centering
    \includegraphics[width=0.48\textwidth]{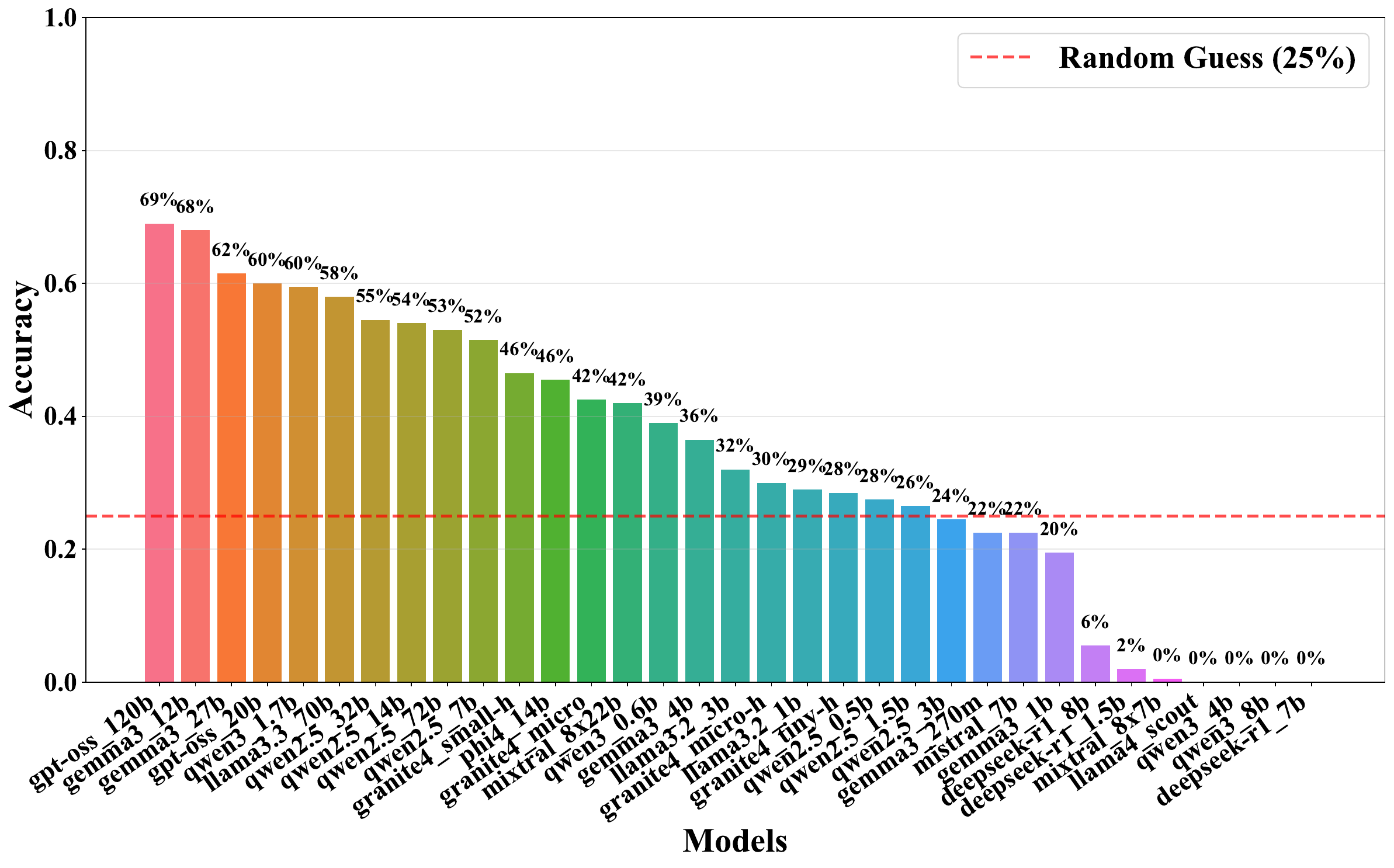}
    \hfill
    \phantom{\includegraphics[width=0.48\textwidth]{figs/bit_vis/8_bit_comparison.pdf}}
    \caption{Accuracy ranking for the 8-qubit benchmark. Only a handful of data-verified copilots (notably \texttt{gpt-oss:120b}, \texttt{gemma3:27b}, \texttt{qwen2.5:32b/72b}) stay above 60\%, whereas the majority of general-purpose checkpoints fall below 30\% or flatline at zero, underscoring how qubit scaling amplifies the need for external verification.}
    \label{fig:bit-row4}
\end{figure*}

\begin{figure*}[t]
    \centering
    \includegraphics[width=0.32\textwidth]{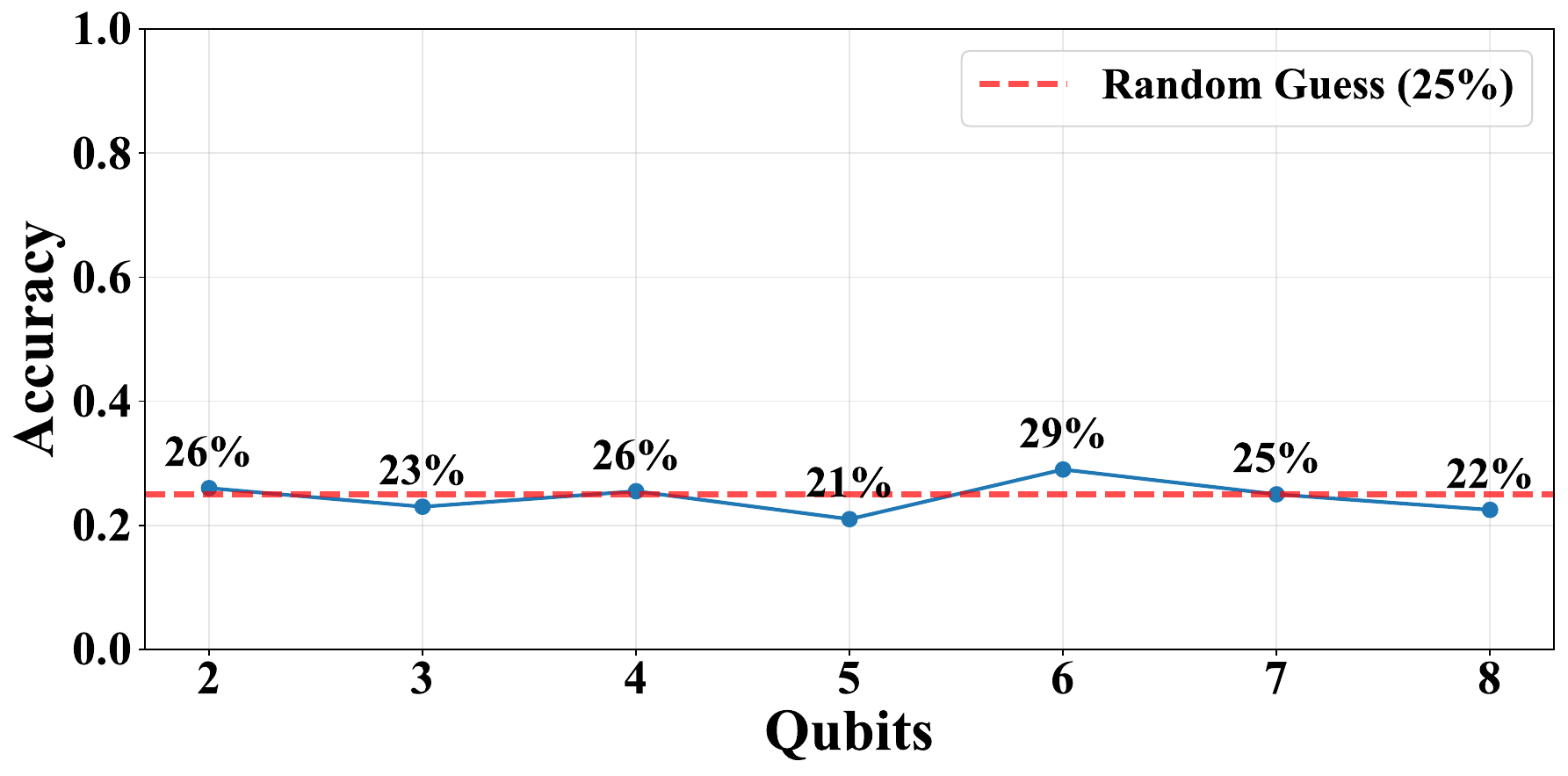}
    \includegraphics[width=0.32\textwidth]{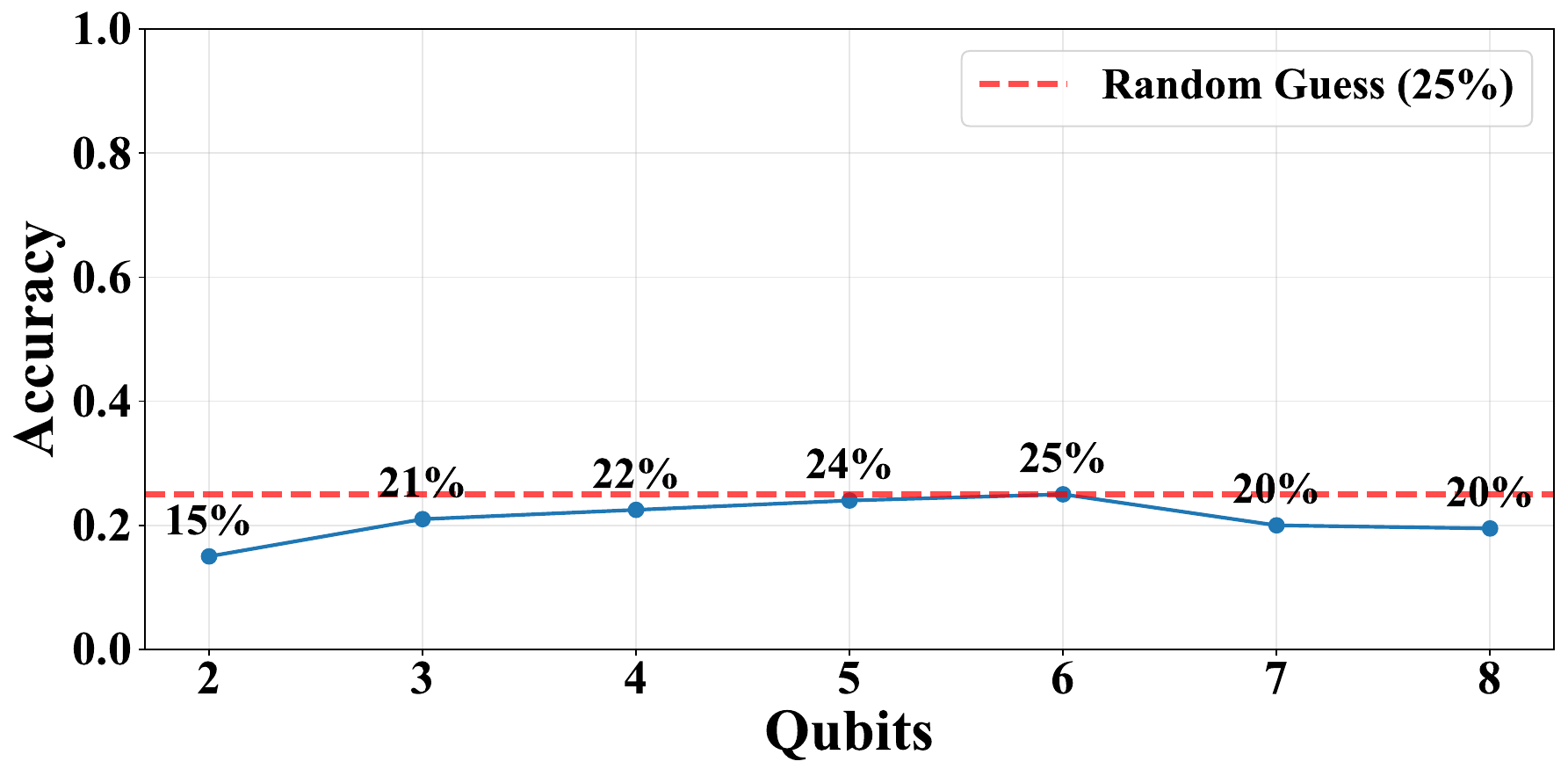}
    \includegraphics[width=0.32\textwidth]{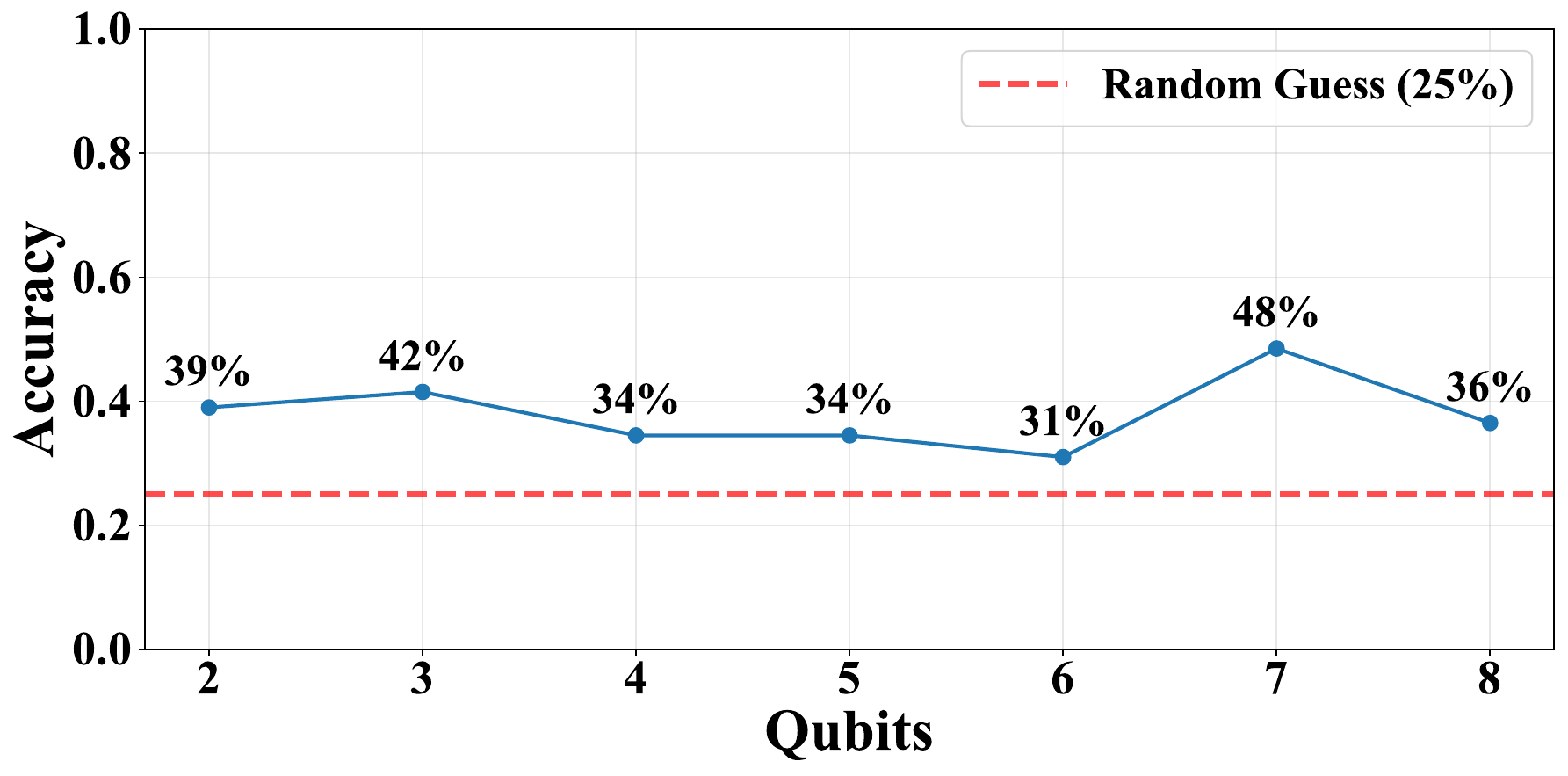}
    \caption{Performance of the \texttt{gemma3:270m}, \texttt{gemma3:1b}, and \texttt{gemma3:4b} checkpoints trained only with generic instruction data. All three curves hug the 25\% random baseline—\texttt{gemma3:4b} briefly reaches $\sim$40\% on 2--3 qubits before decaying toward chance beyond six—showing that parameter count alone does not recover MAJ/UMA reasoning.}
    \label{fig:model-row1}
\end{figure*}

\begin{figure*}[t]
    \centering
    \includegraphics[width=0.32\textwidth]{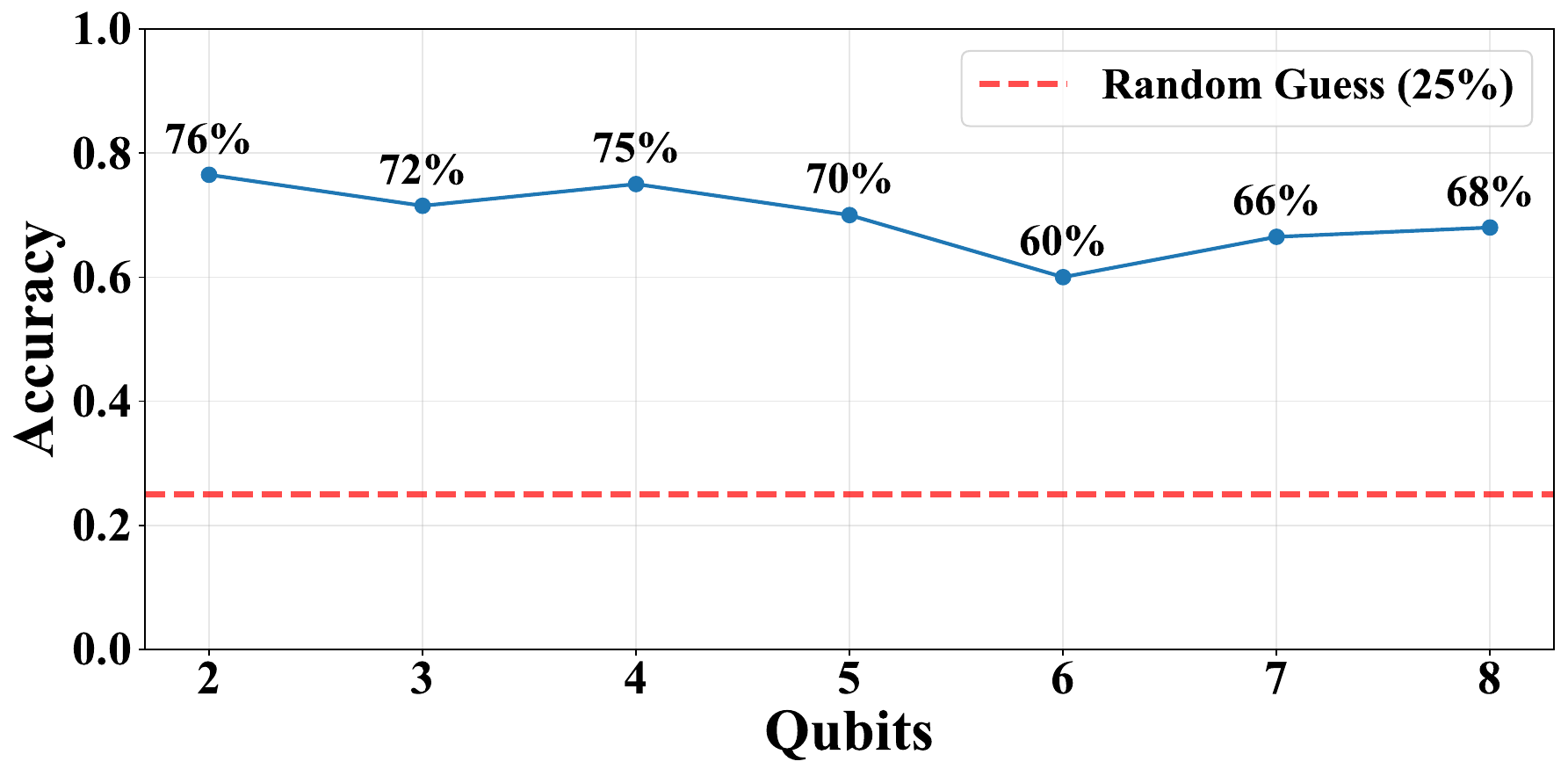}
    \includegraphics[width=0.32\textwidth]{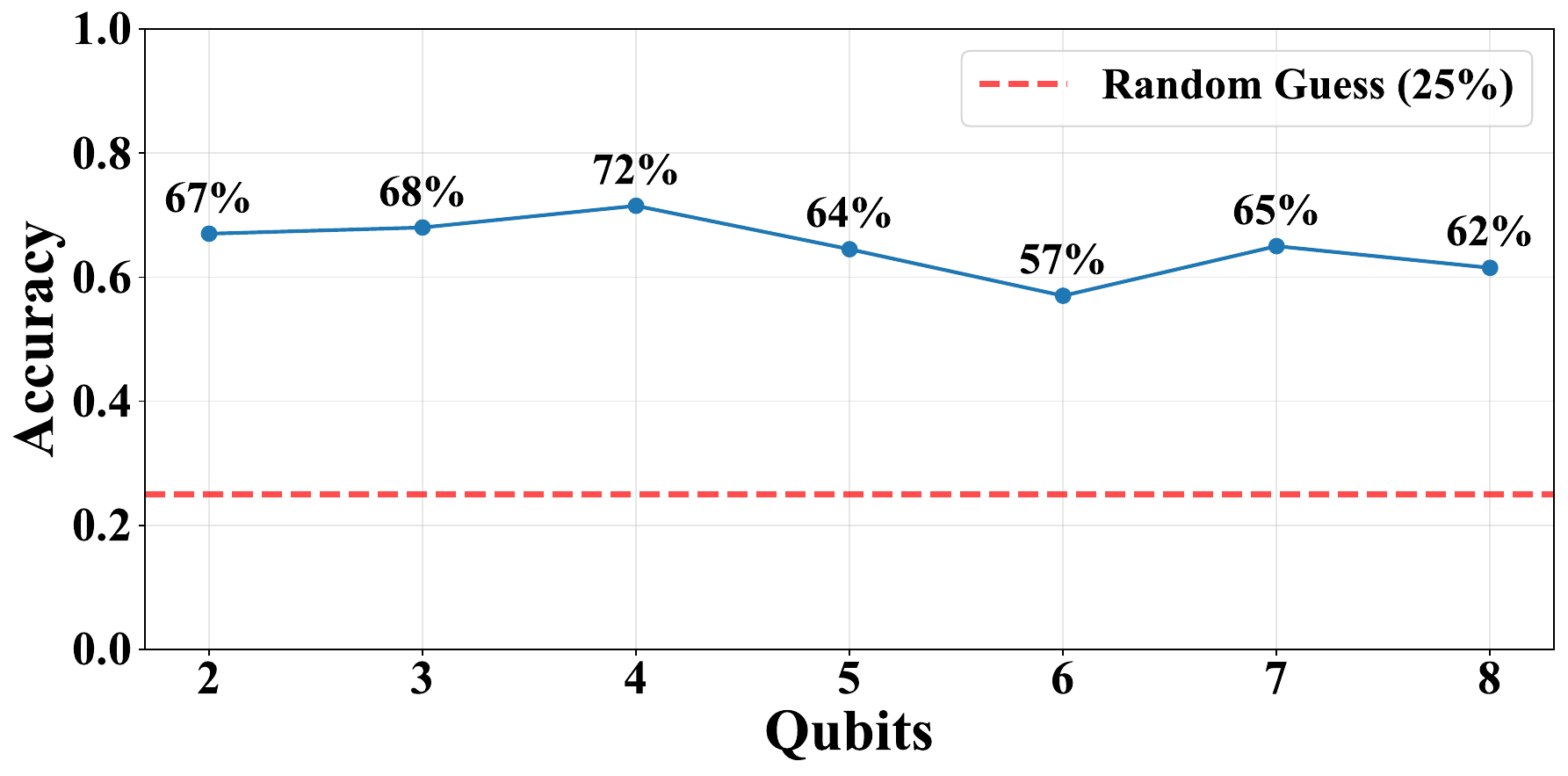}
    \includegraphics[width=0.32\textwidth]{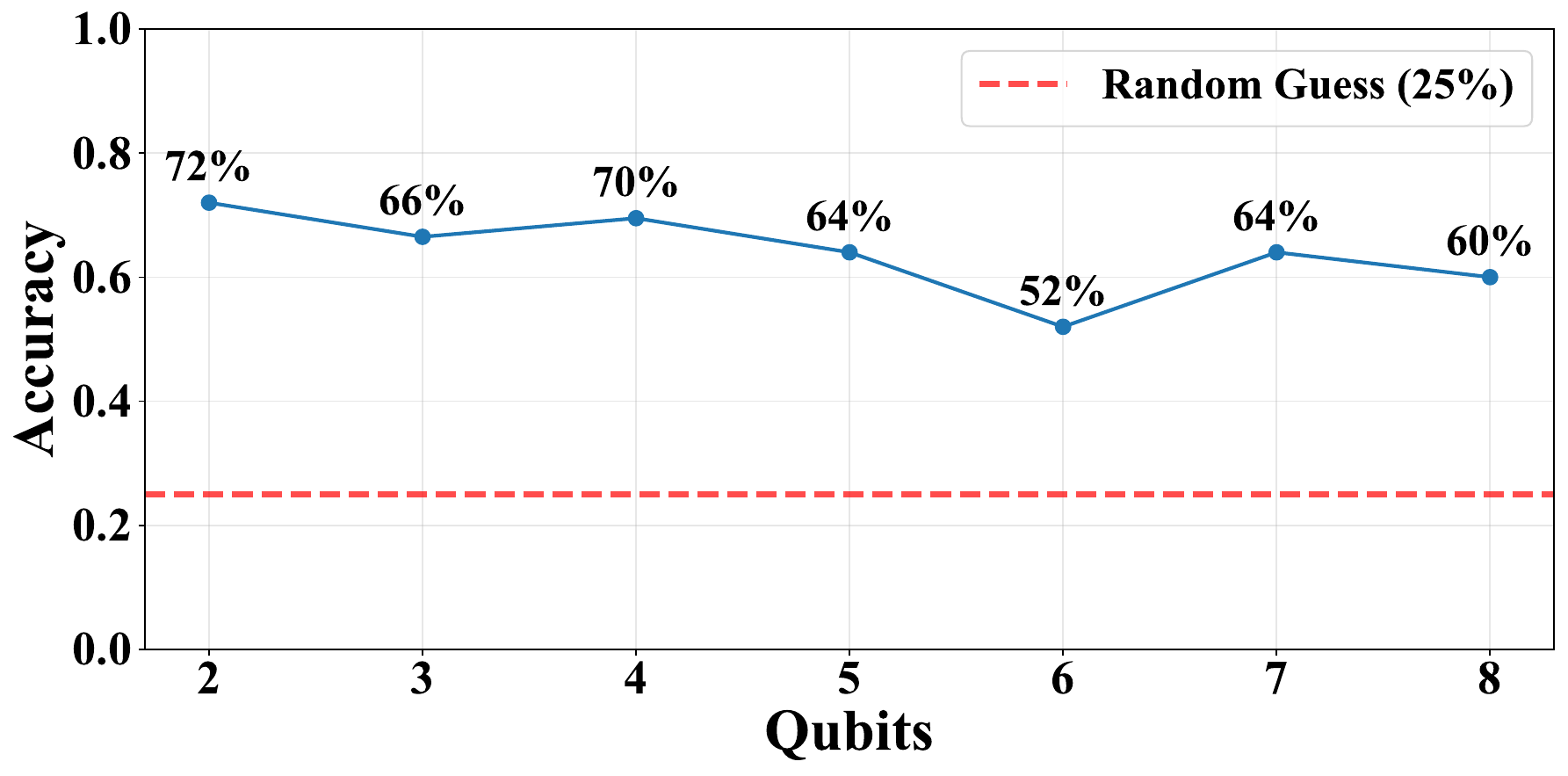}
    \caption{Verification-conditioned \texttt{gemma3:12b}, \texttt{gemma3:27b}, and \texttt{gpt-oss:20b}. These checkpoints hold 0.60--0.75 accuracy through all qubit counts (with \texttt{gpt-oss:20b} dipping only to 0.52 at 8 qubits), illustrating the large jump gained from verified data compared with Figure~\ref{fig:model-row1}.}
    \label{fig:model-row2}
\end{figure*}

\begin{figure*}[t]
    \centering
    \includegraphics[width=0.32\textwidth]{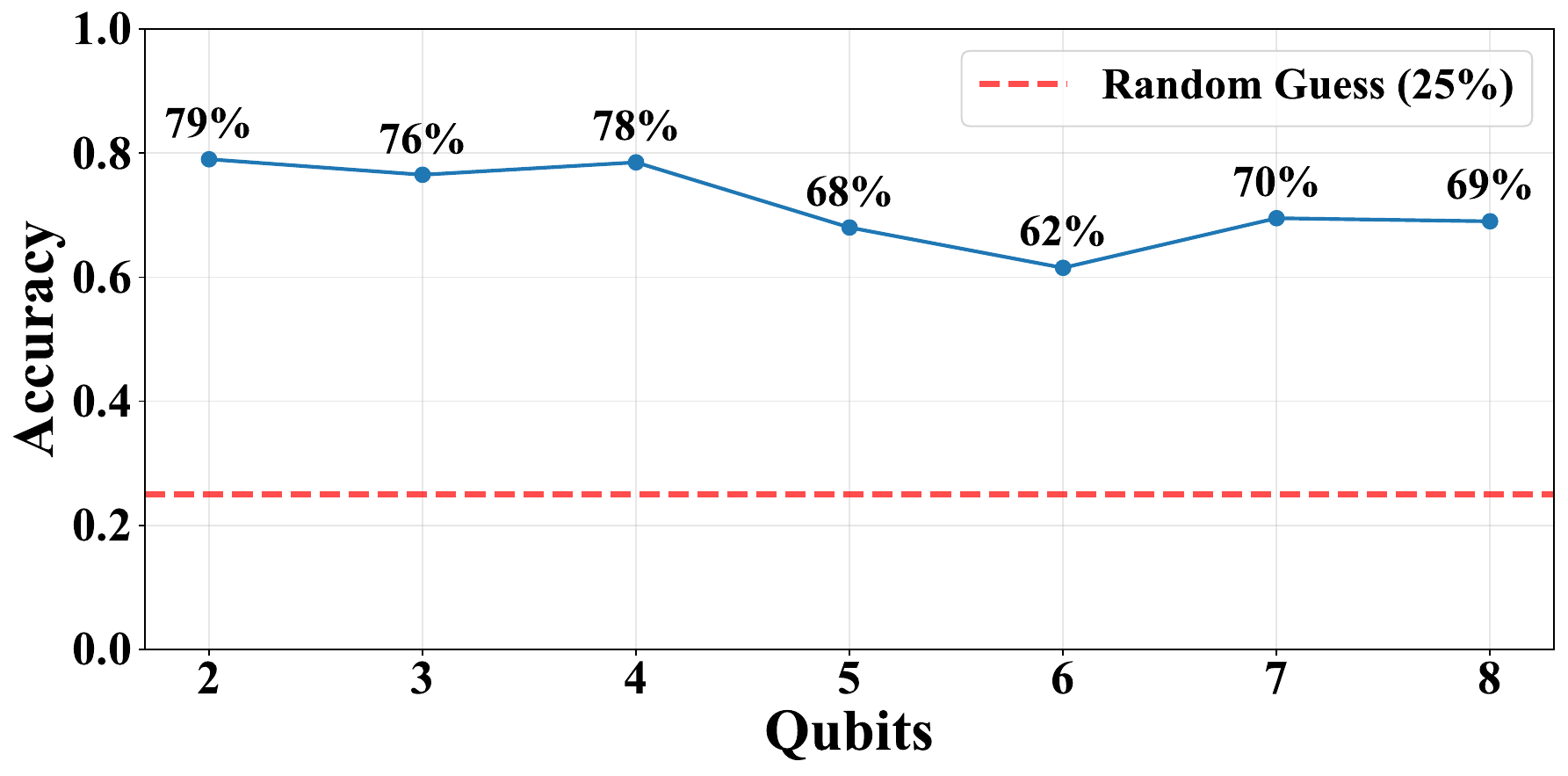}
    \includegraphics[width=0.32\textwidth]{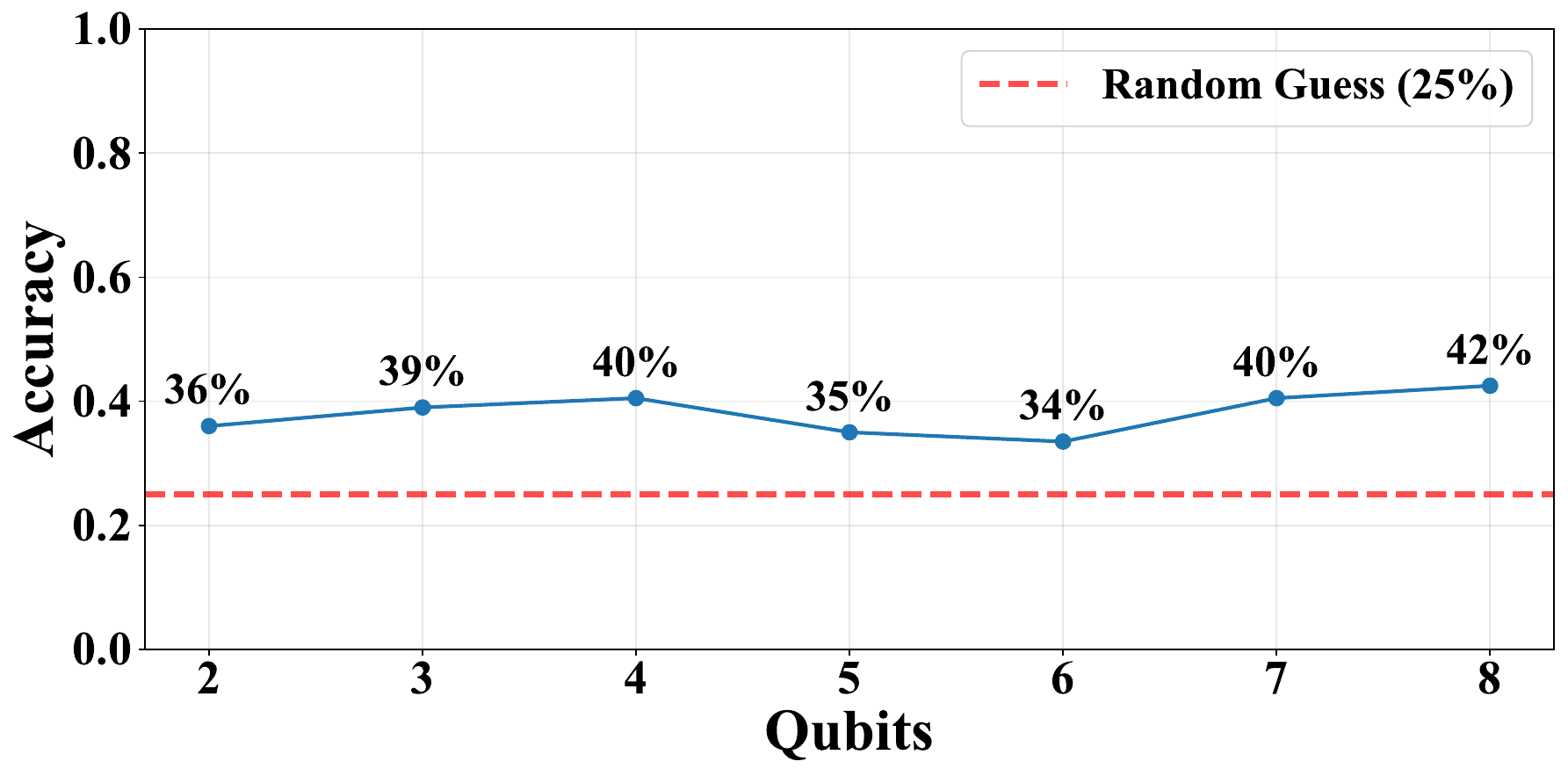}
    \includegraphics[width=0.32\textwidth]{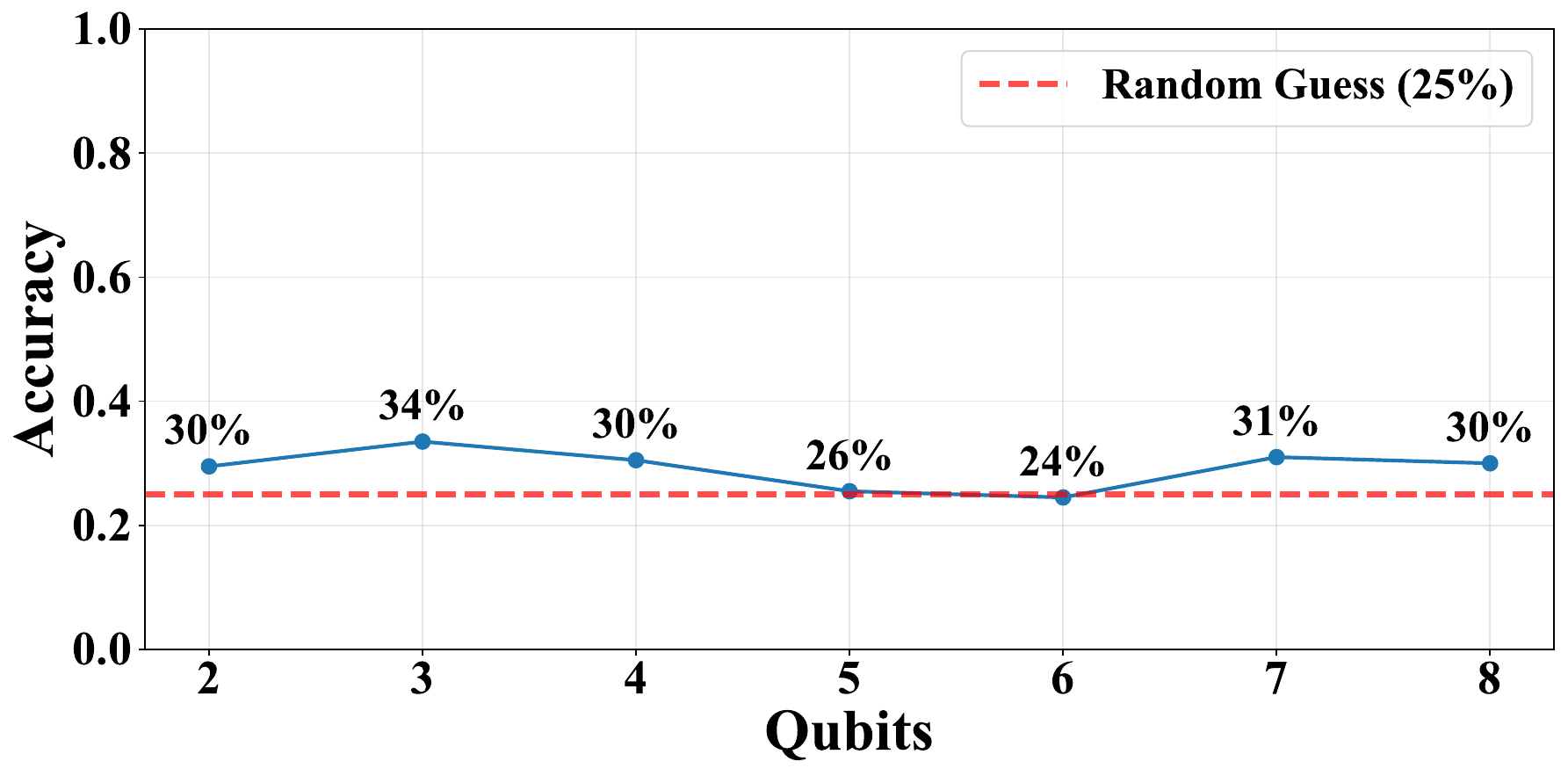}
    \caption{Comparison between \texttt{gpt-oss:120b} and the compact \texttt{granite4} copilots. The 120B model stays between 0.62 and 0.79 accuracy even at 8 qubits, whereas \texttt{granite4:micro} and \texttt{granite4:micro-h} fluctuate in the 0.30--0.42 band and erode with circuit depth, highlighting the gap between verification-first and distilled instruction-tuning.}
    \label{fig:model-row3}
\end{figure*}

\begin{figure*}[t]
    \centering
    \includegraphics[width=0.32\textwidth]{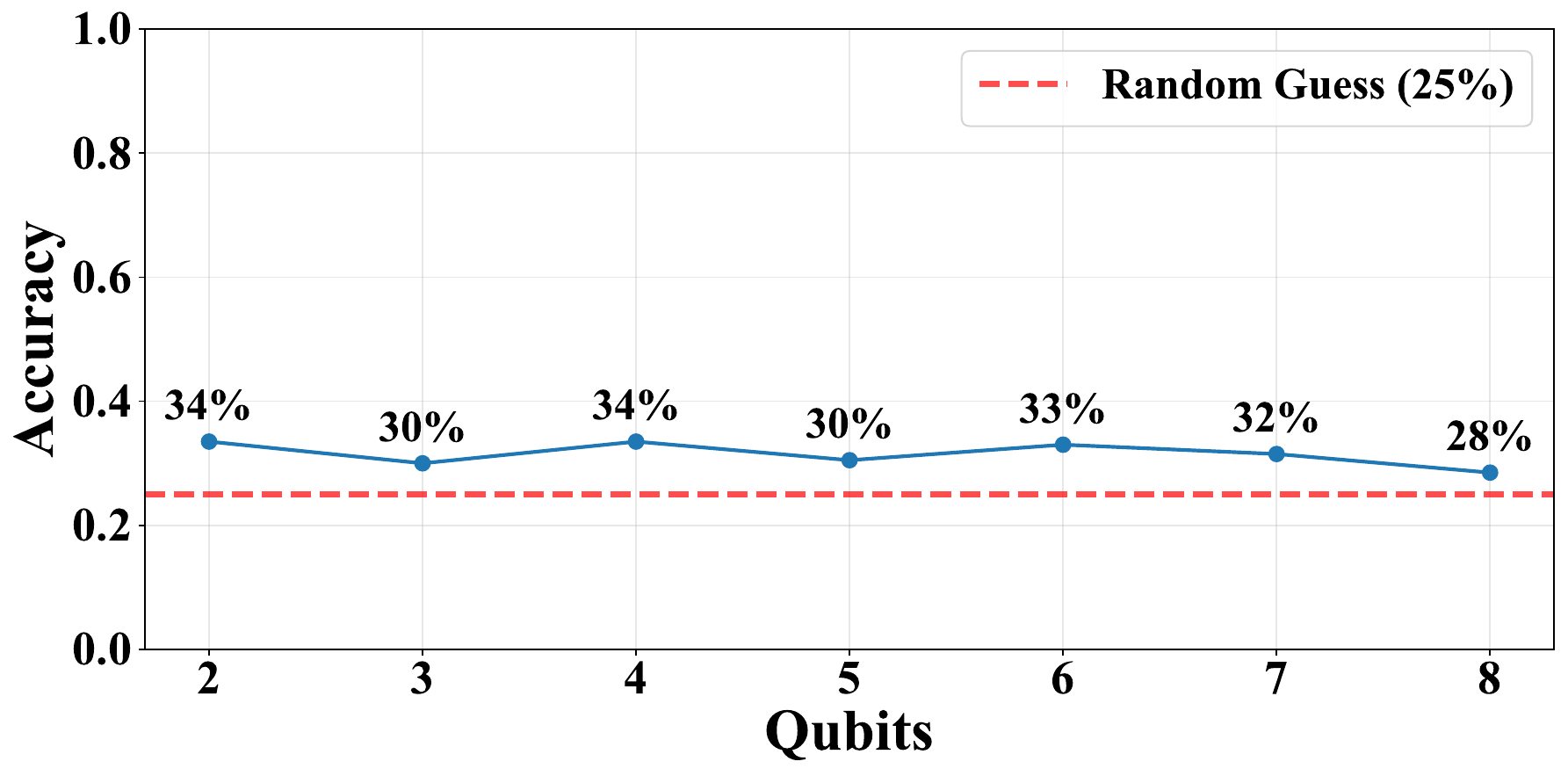}
    \includegraphics[width=0.32\textwidth]{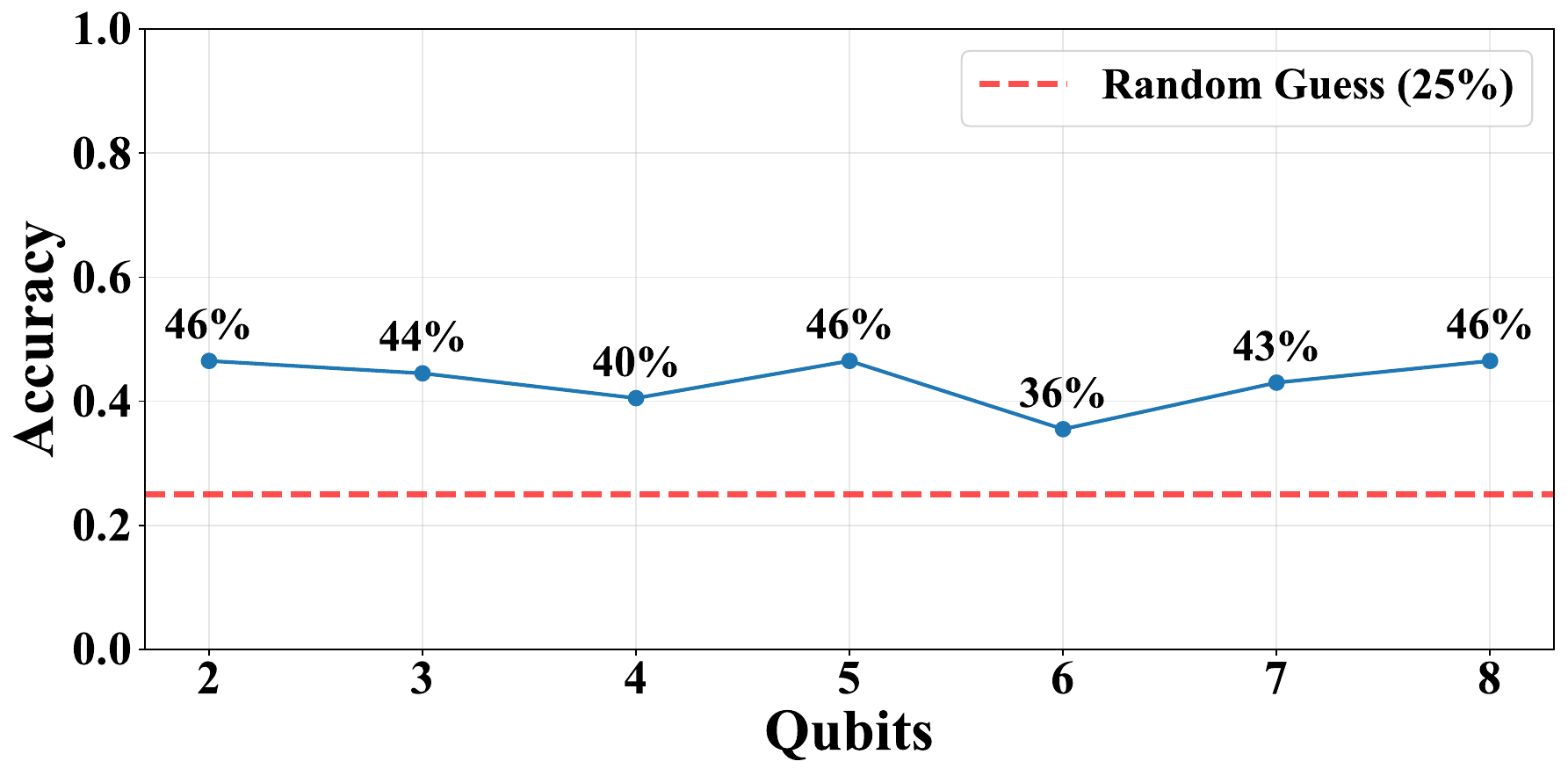}
    \includegraphics[width=0.32\textwidth]{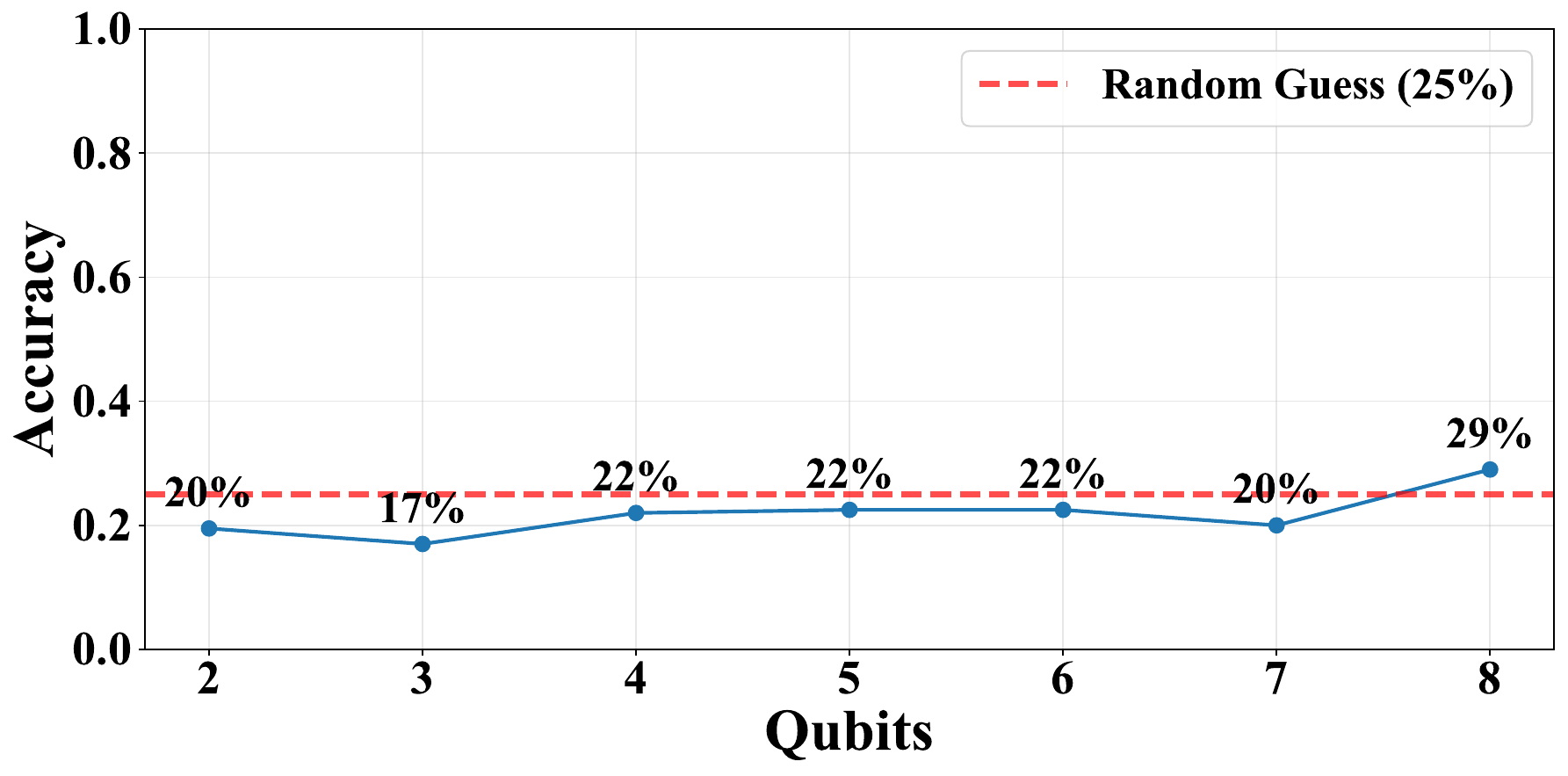}
    \caption{Effect of lighter \texttt{granite4} releases versus a 1B Llama. \texttt{granite4:tiny-h} and \texttt{granite4:small-h} hover in the low 30--40\% range without scaling benefits, and \texttt{llama3.2:1b} drops from 0.29 to $\approx$0.20 as qubit counts grow, reinforcing that lightweight copilots need explicit verification signals.}
    \label{fig:model-row4}
\end{figure*}

\begin{figure*}[t]
    \centering
    \includegraphics[width=0.32\textwidth]{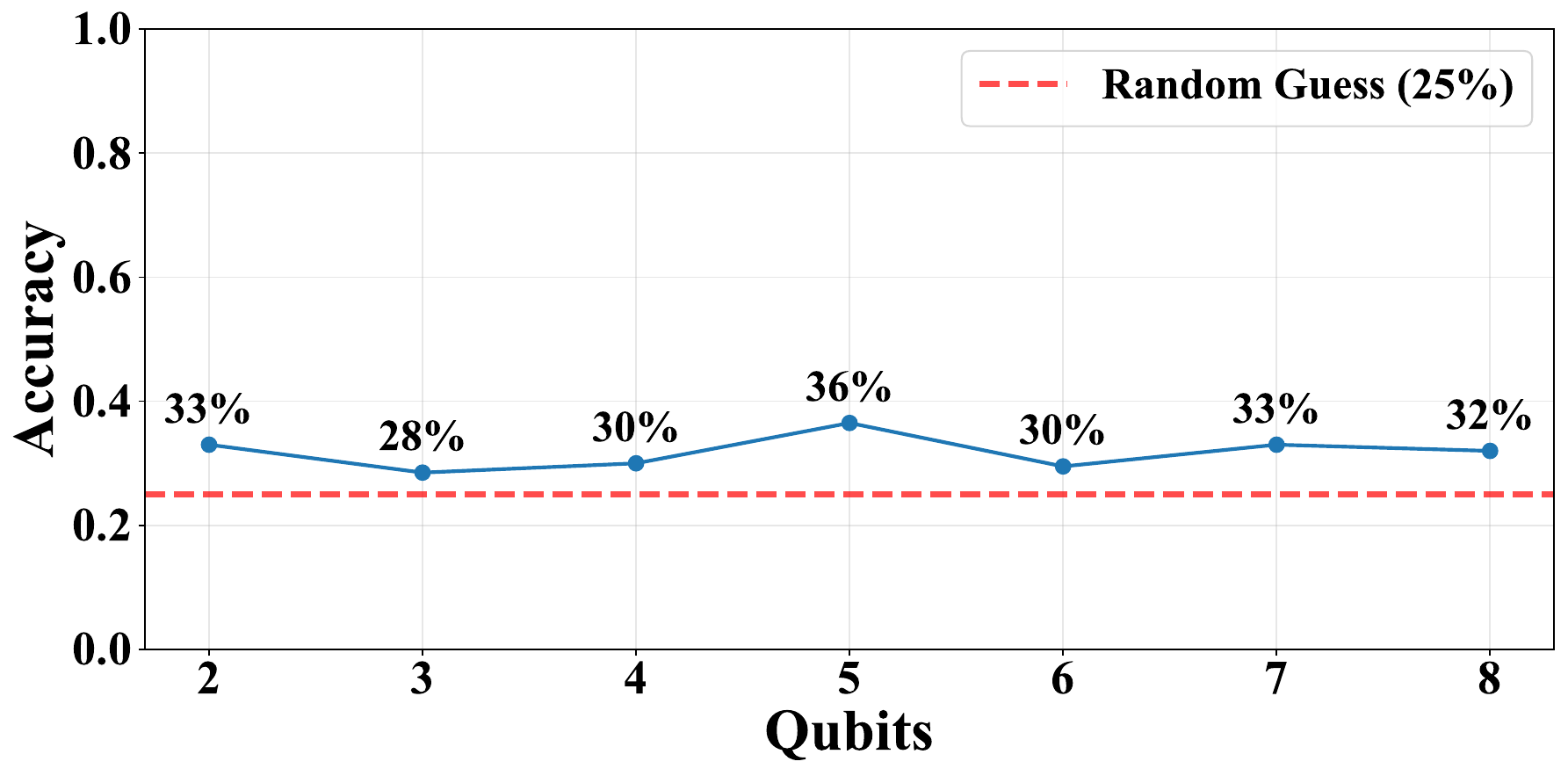}
    \includegraphics[width=0.32\textwidth]{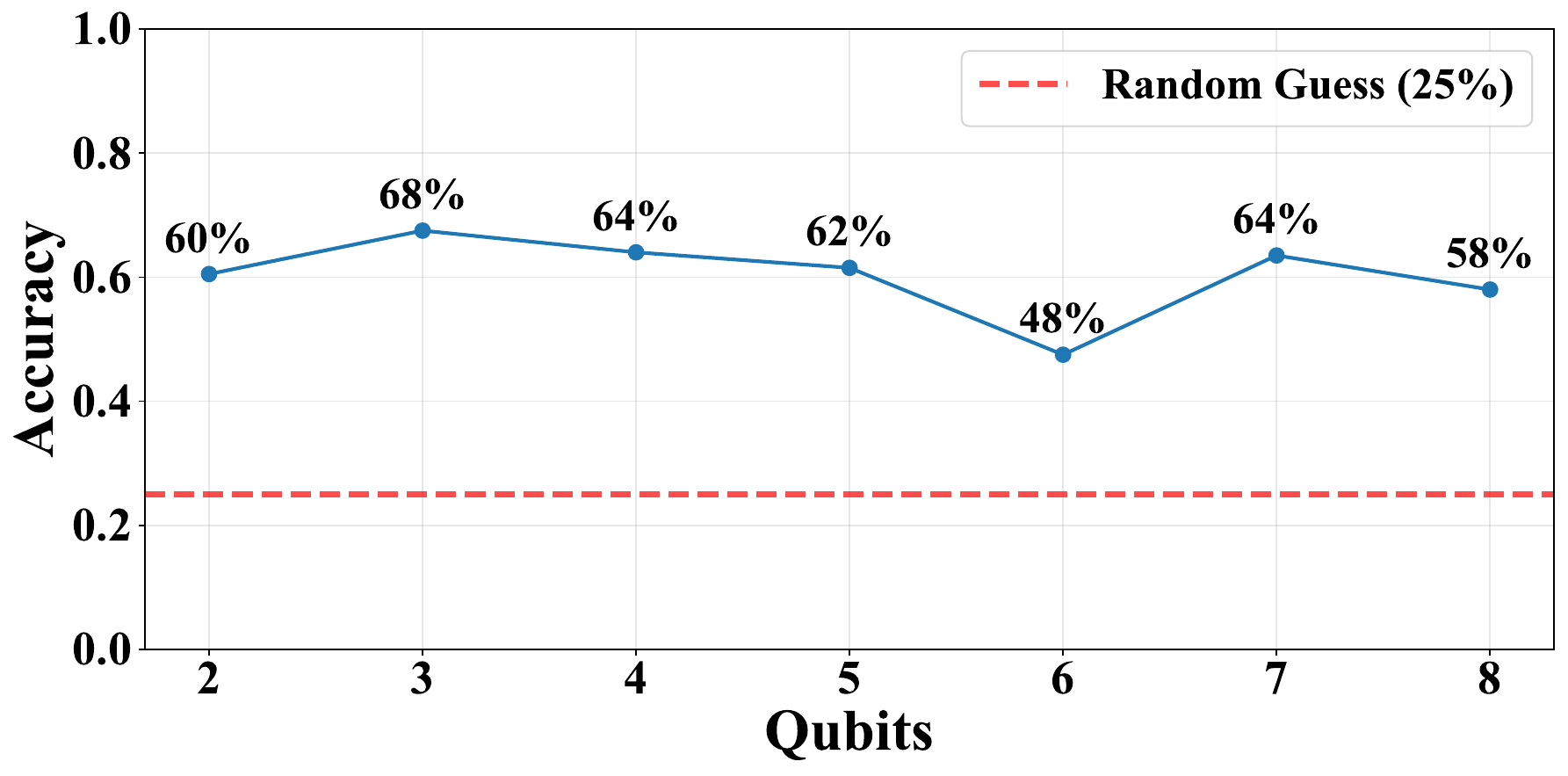}
    \includegraphics[width=0.32\textwidth]{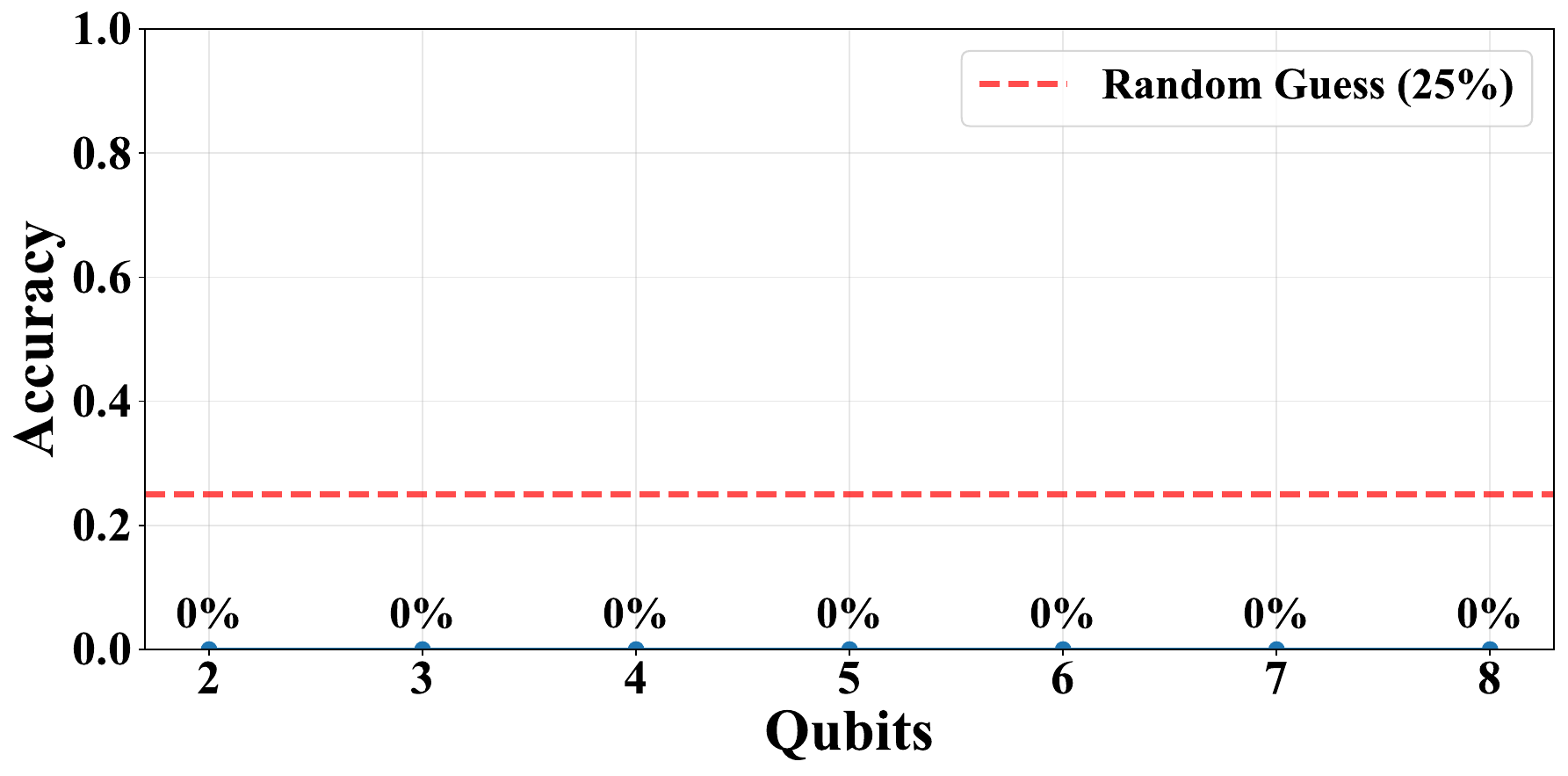}
    \caption{Llama family scaling. \texttt{llama3.2:3b} remains near 0.30 accuracy across qubits, \texttt{llama3.3:70b} peaks at 0.64--0.68 on 2--5 qubits before tapering to 0.48 by 8 qubits, and the distilled \texttt{llama4:scout} agent fails to answer any instance correctly, underscoring inconsistent gains from size alone.}
    \label{fig:model-row5}
\end{figure*}

\begin{figure*}[t]
    \centering
    \includegraphics[width=0.32\textwidth]{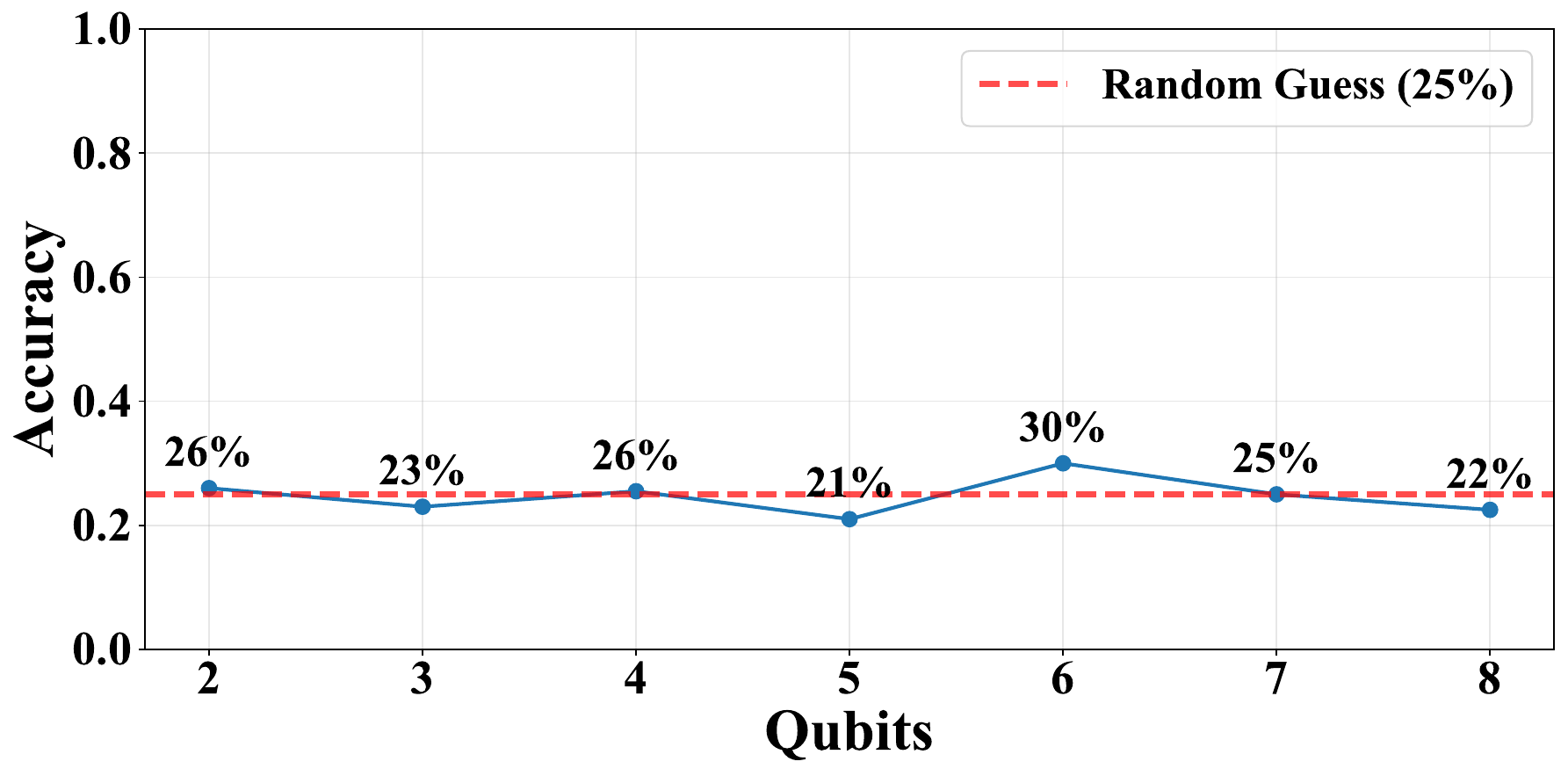}
    \includegraphics[width=0.32\textwidth]{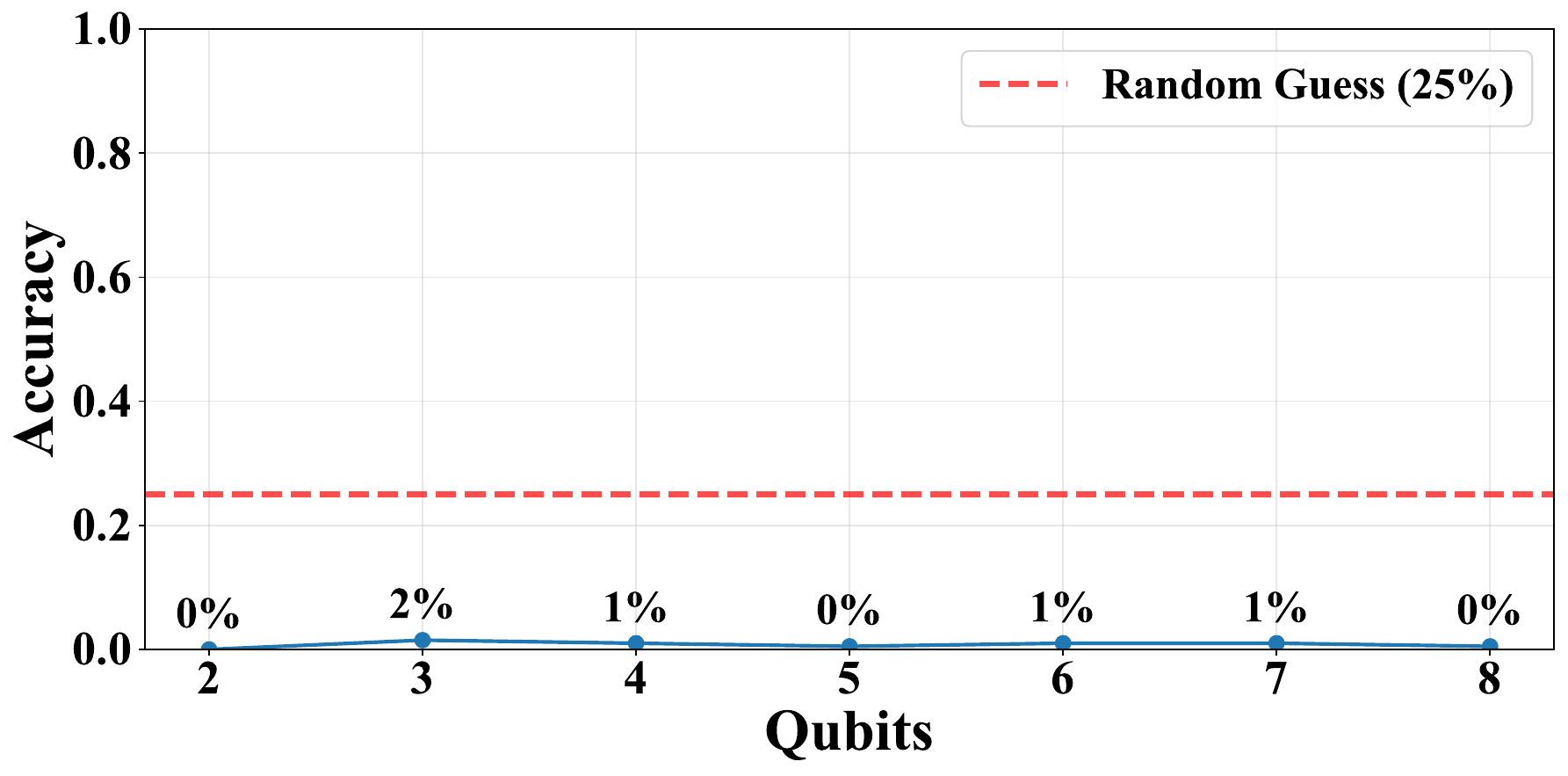}
    \includegraphics[width=0.32\textwidth]{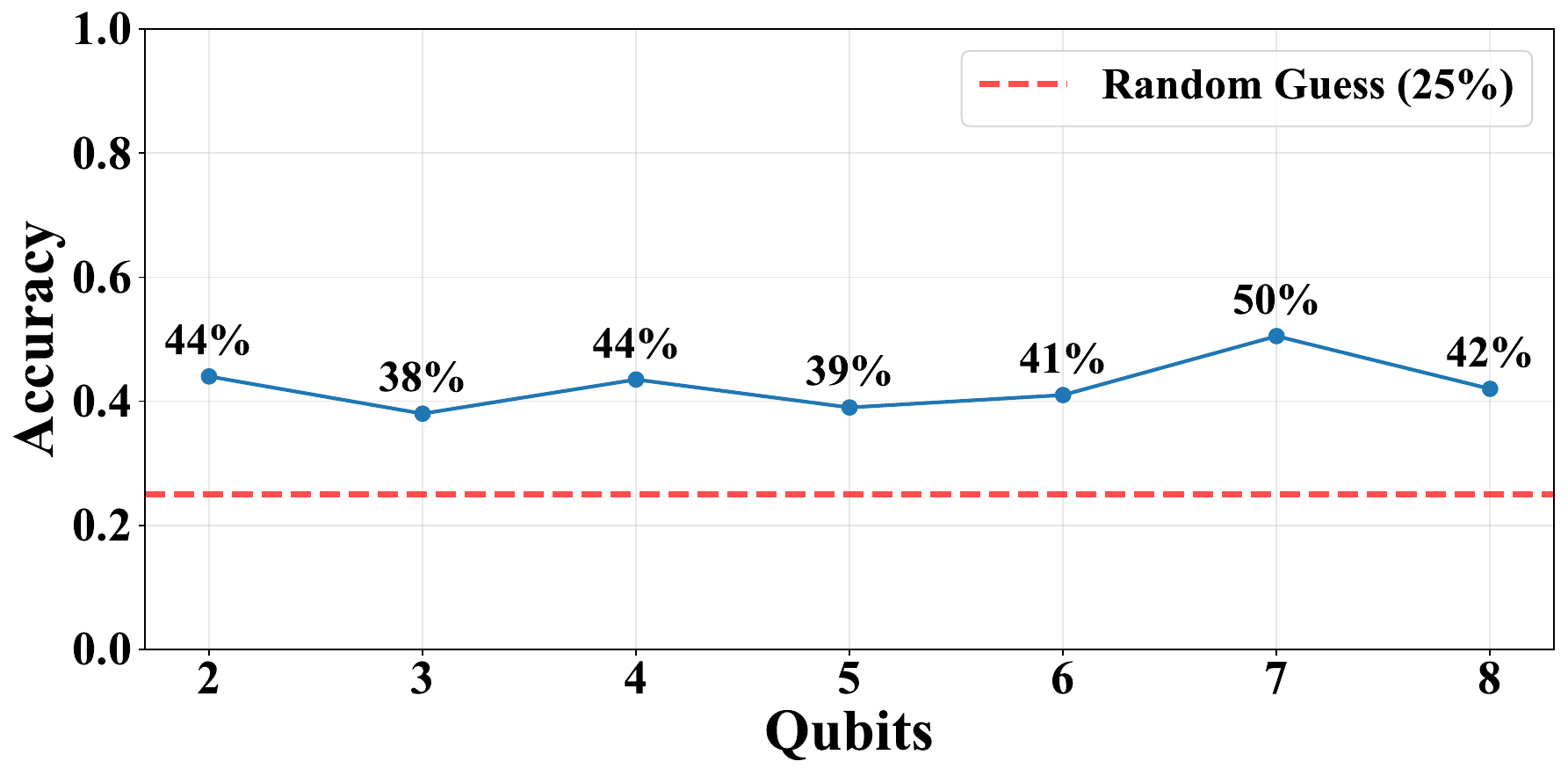}
    \caption{Mistral and Mixtral checkpoints. The dense \texttt{mistral:7b} oscillates around 0.25 accuracy, the sparse \texttt{mixtral:8x7b} sits at $\leq$2\% on every qubit count, and \texttt{mixtral:8x22b} briefly reaches 0.44--0.50 before sliding back toward the high-30\% range, showing that mixture-of-experts routing without verification does not guarantee reliability.}
    \label{fig:model-row6}
\end{figure*}

\begin{figure*}[t]
    \centering
    \includegraphics[width=0.32\textwidth]{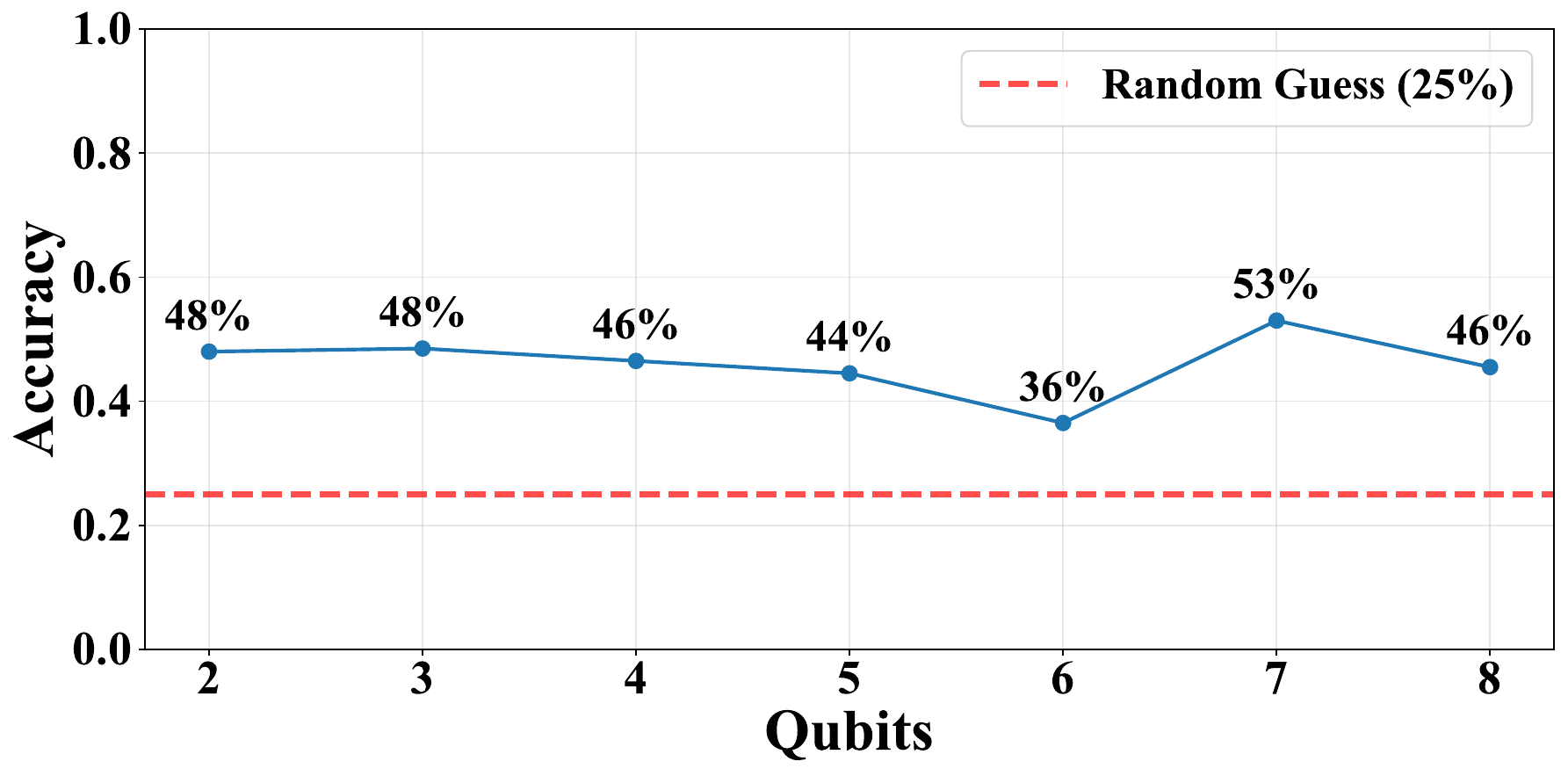}
    \includegraphics[width=0.32\textwidth]{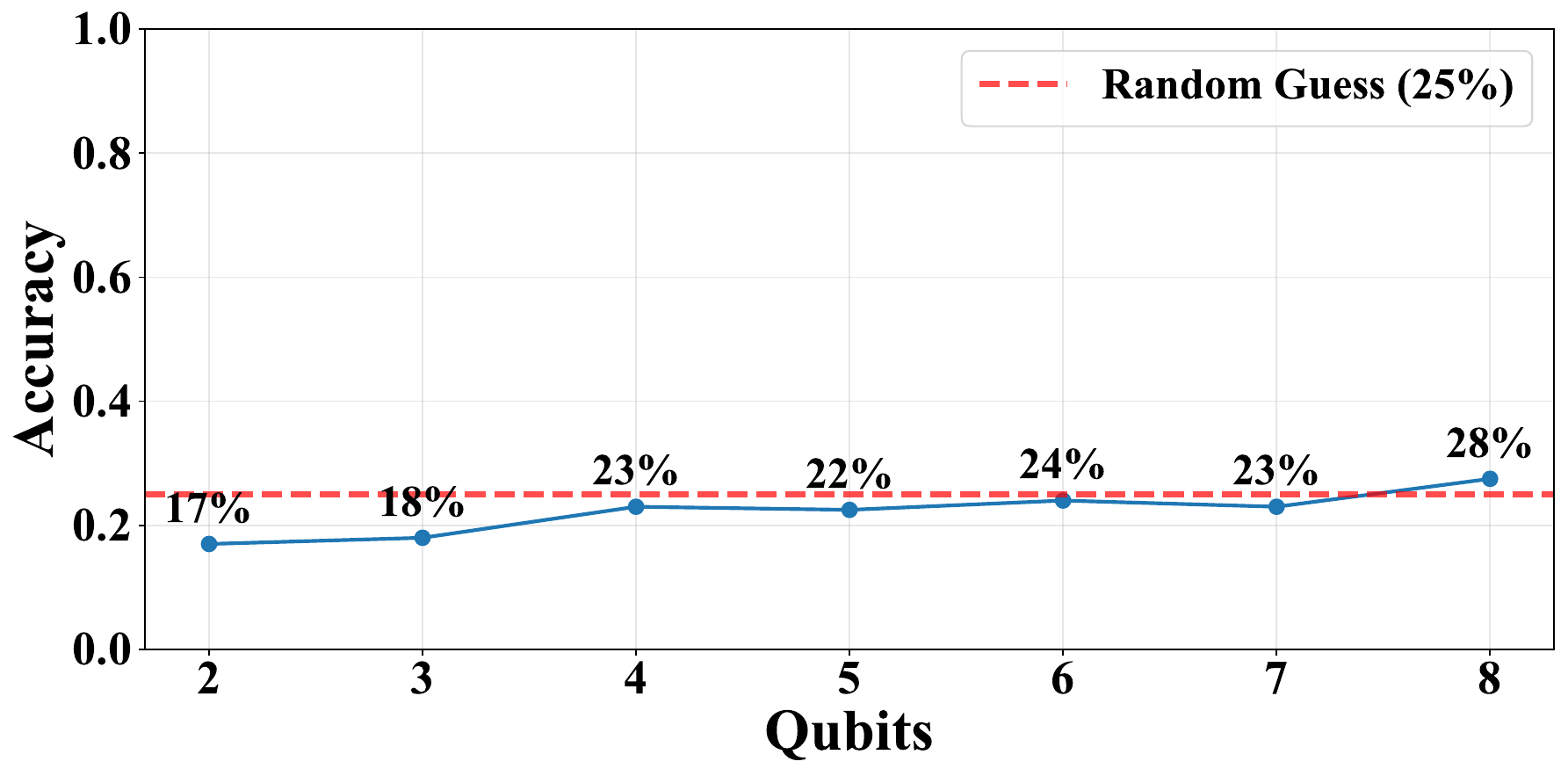}
    \includegraphics[width=0.32\textwidth]{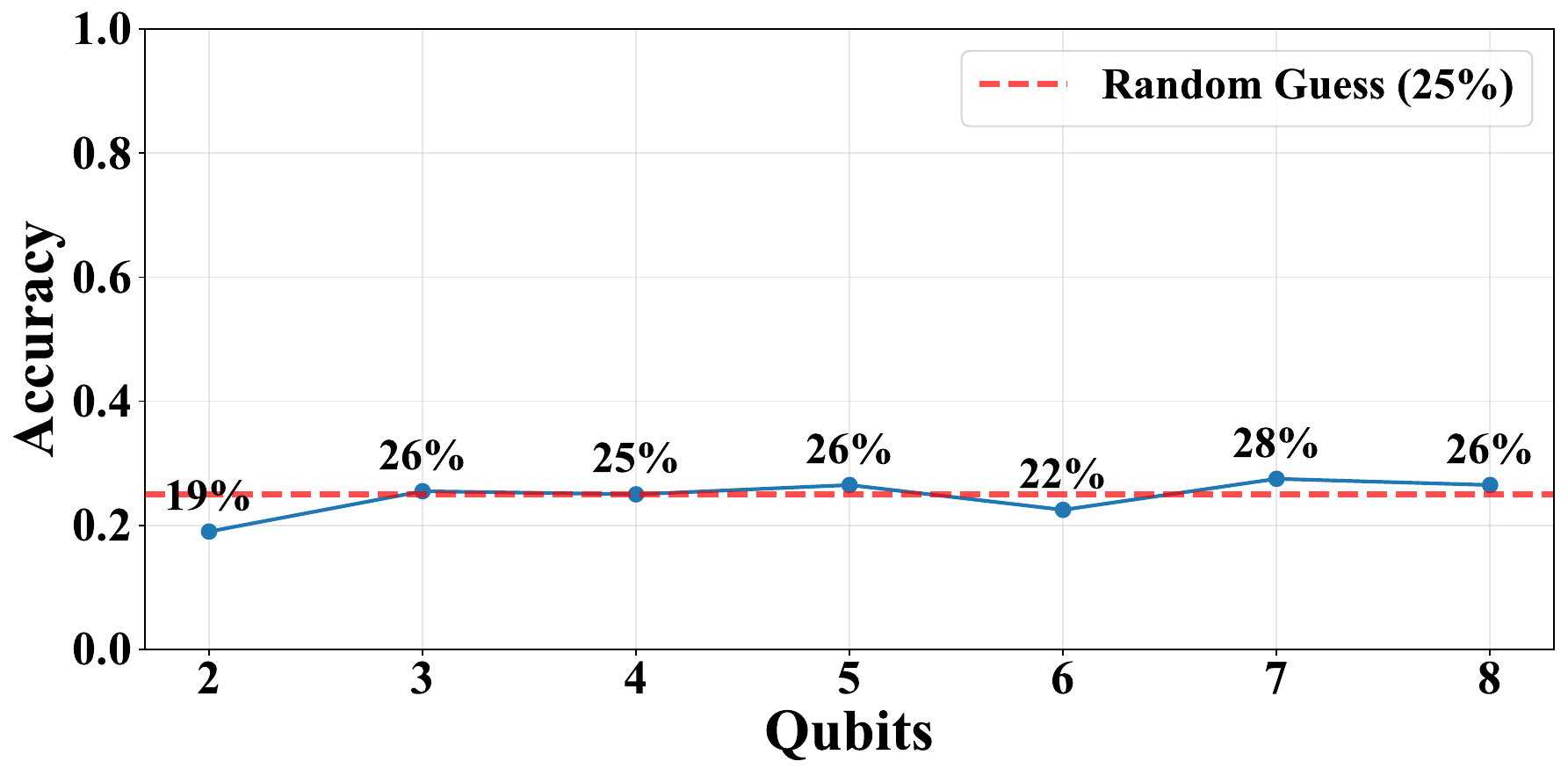}
    \caption{Mid-sized \texttt{phi4} and entry-level \texttt{qwen2.5}. \texttt{phi4:14b} sustains mid-40\% accuracy with a single spike above 50\%, whereas the 0.5B and 1.5B \texttt{qwen2.5} models fluctuate between 0.18 and 0.28 accuracy, barely clearing the random-guess threshold.}
    \label{fig:model-row7}
\end{figure*}

\begin{figure*}[t]
    \centering
    \includegraphics[width=0.32\textwidth]{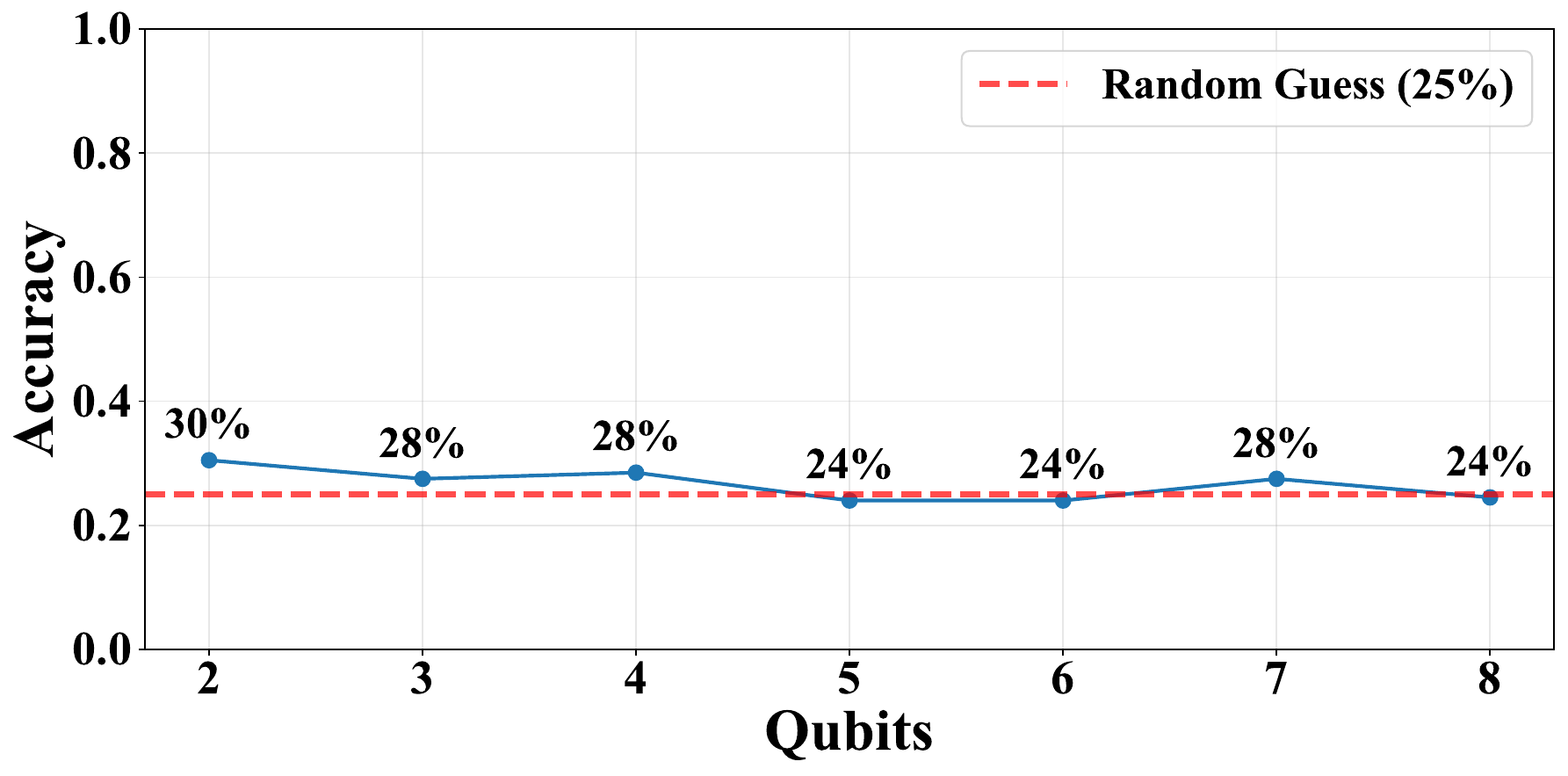}
    \includegraphics[width=0.32\textwidth]{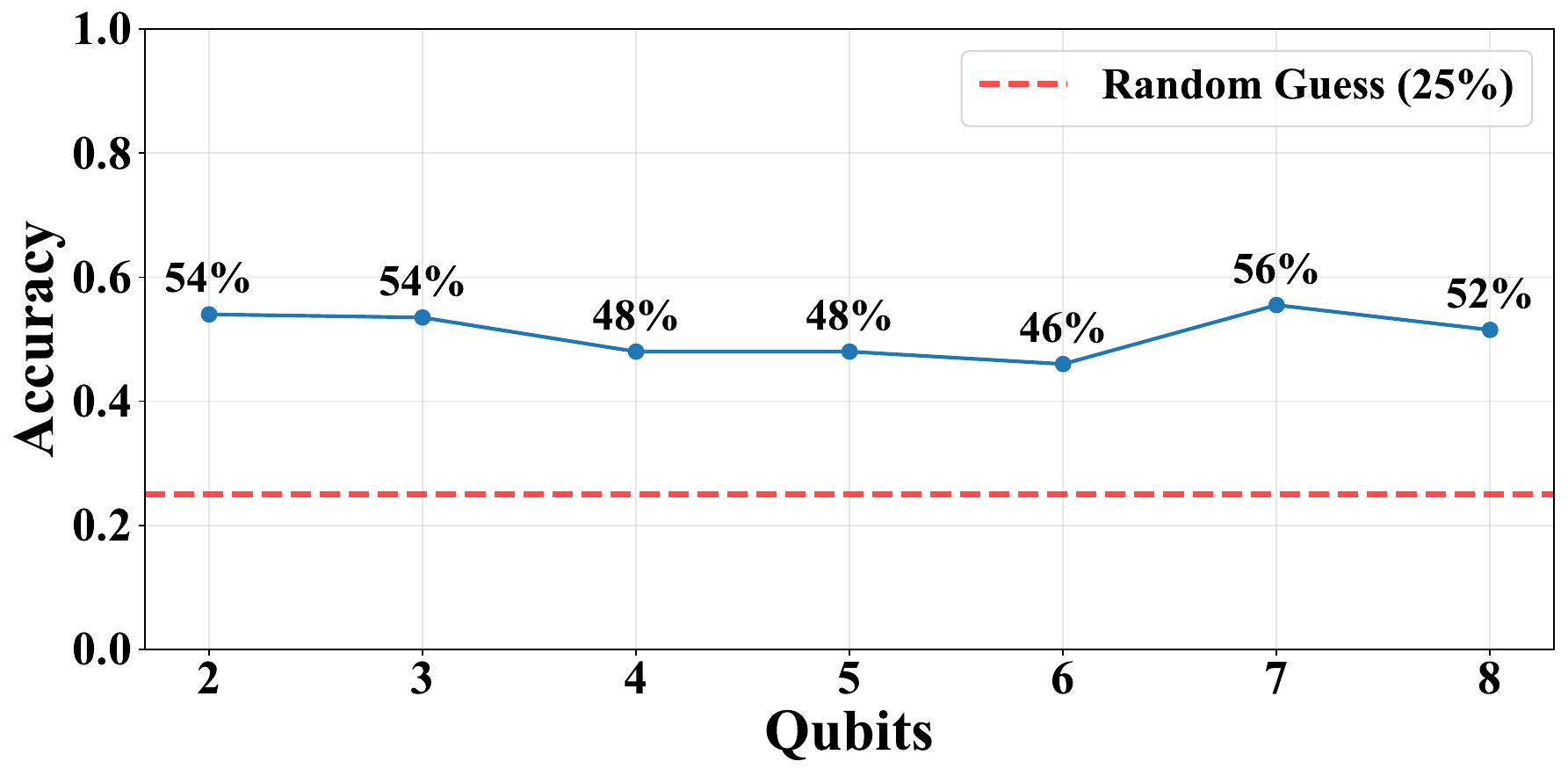}
    \includegraphics[width=0.32\textwidth]{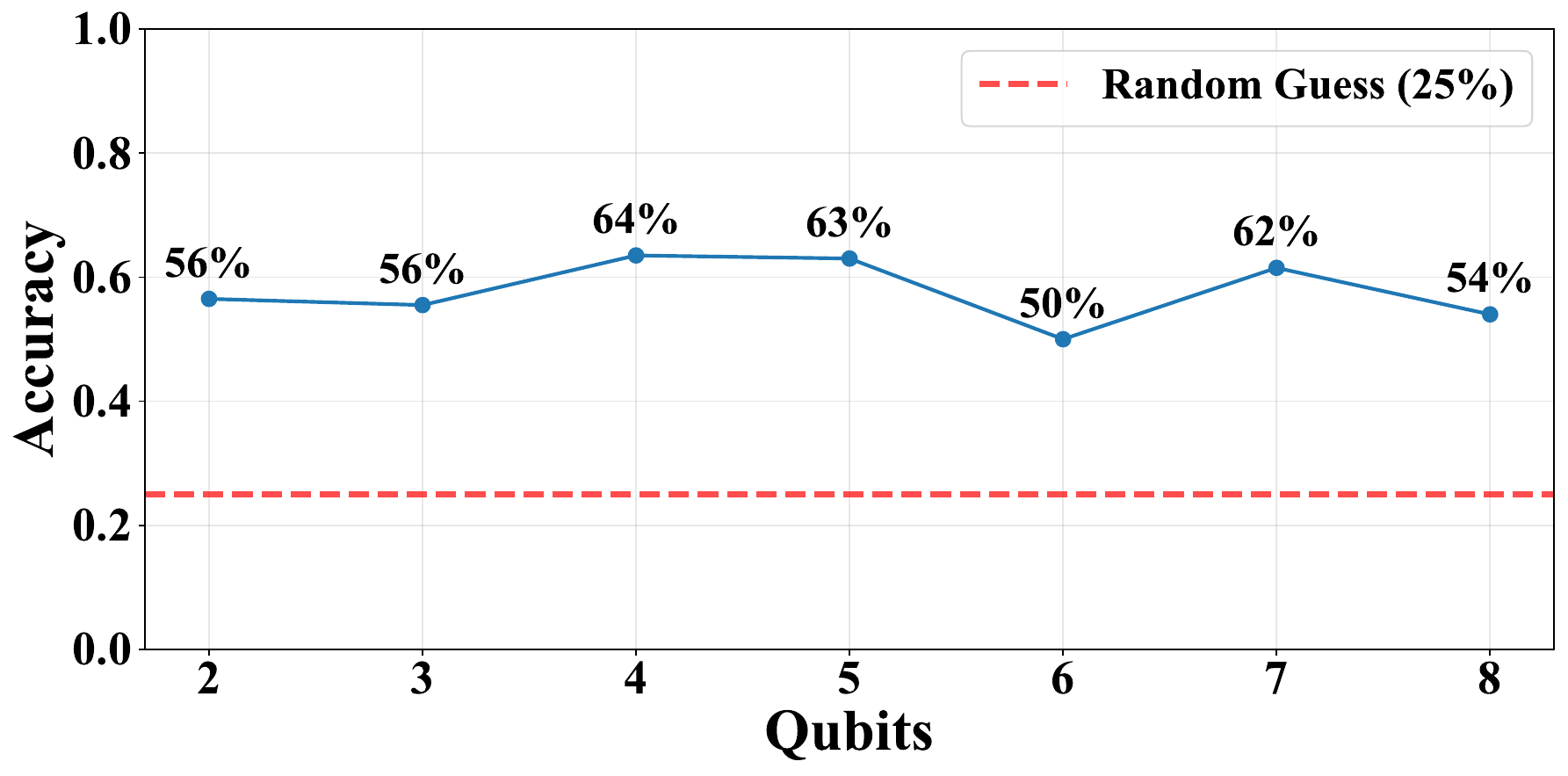}
    \caption{Scaling within \texttt{qwen2.5}. Moving from 3B to 7B to 14B steadily lifts accuracy from $\approx$0.25 to 0.45--0.56 and finally 0.50--0.64, demonstrating that successive parameter doublings under the same training recipe yield material gains on MAJ/UMA verification.}
    \label{fig:model-row8}
\end{figure*}

\begin{figure*}[t]
    \centering
    \includegraphics[width=0.32\textwidth]{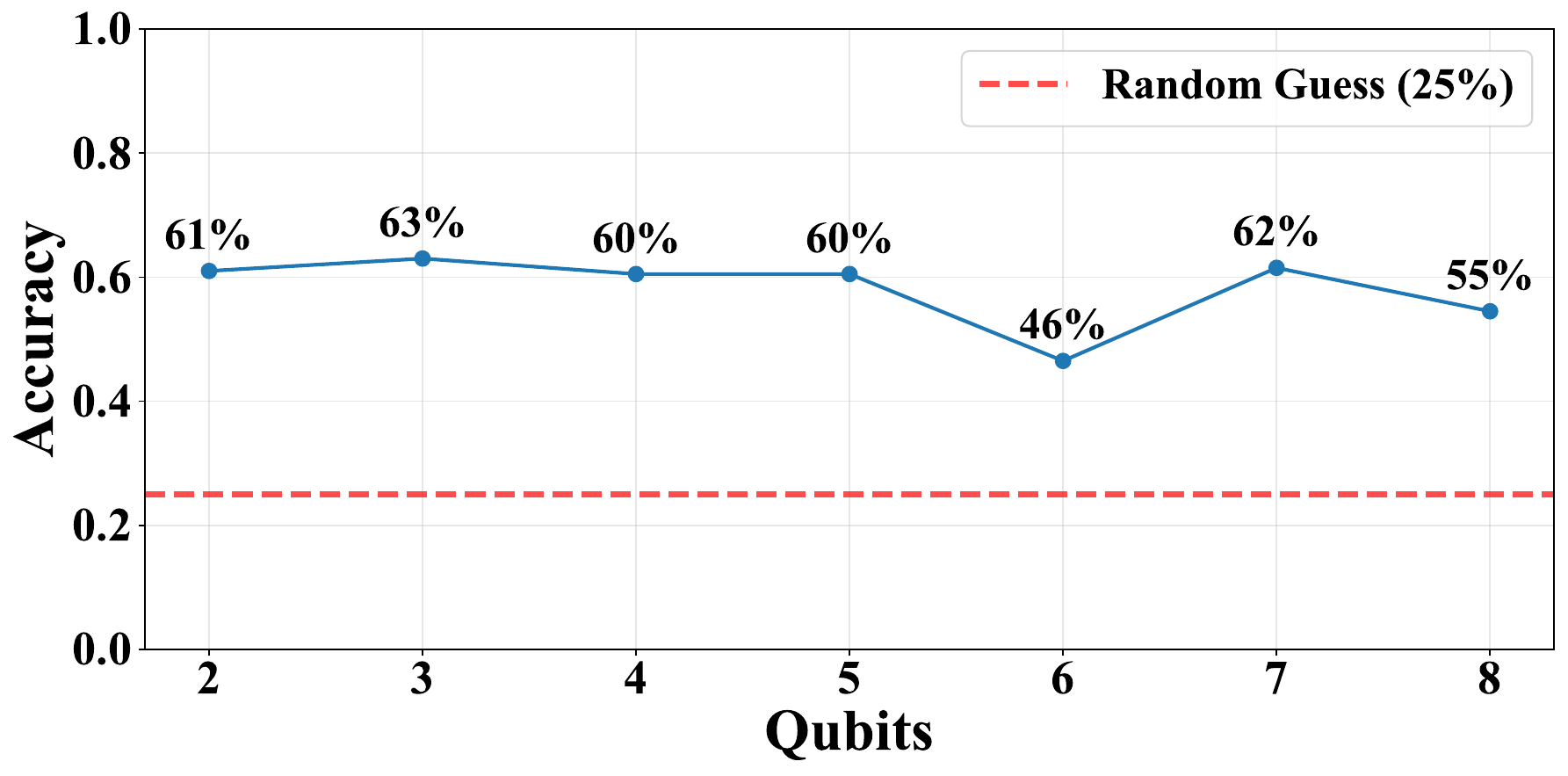}
    \includegraphics[width=0.32\textwidth]{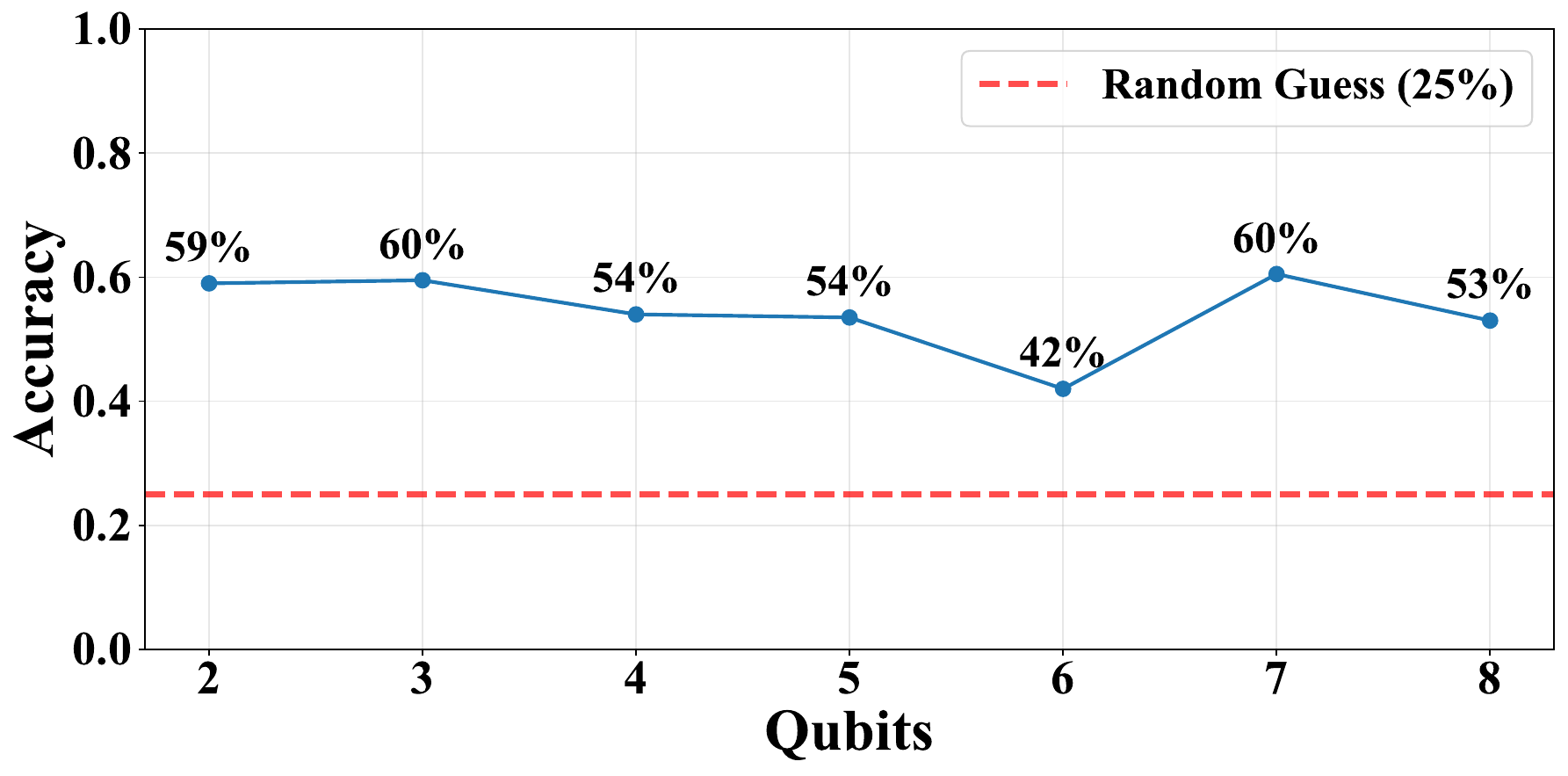}
    \includegraphics[width=0.32\textwidth]{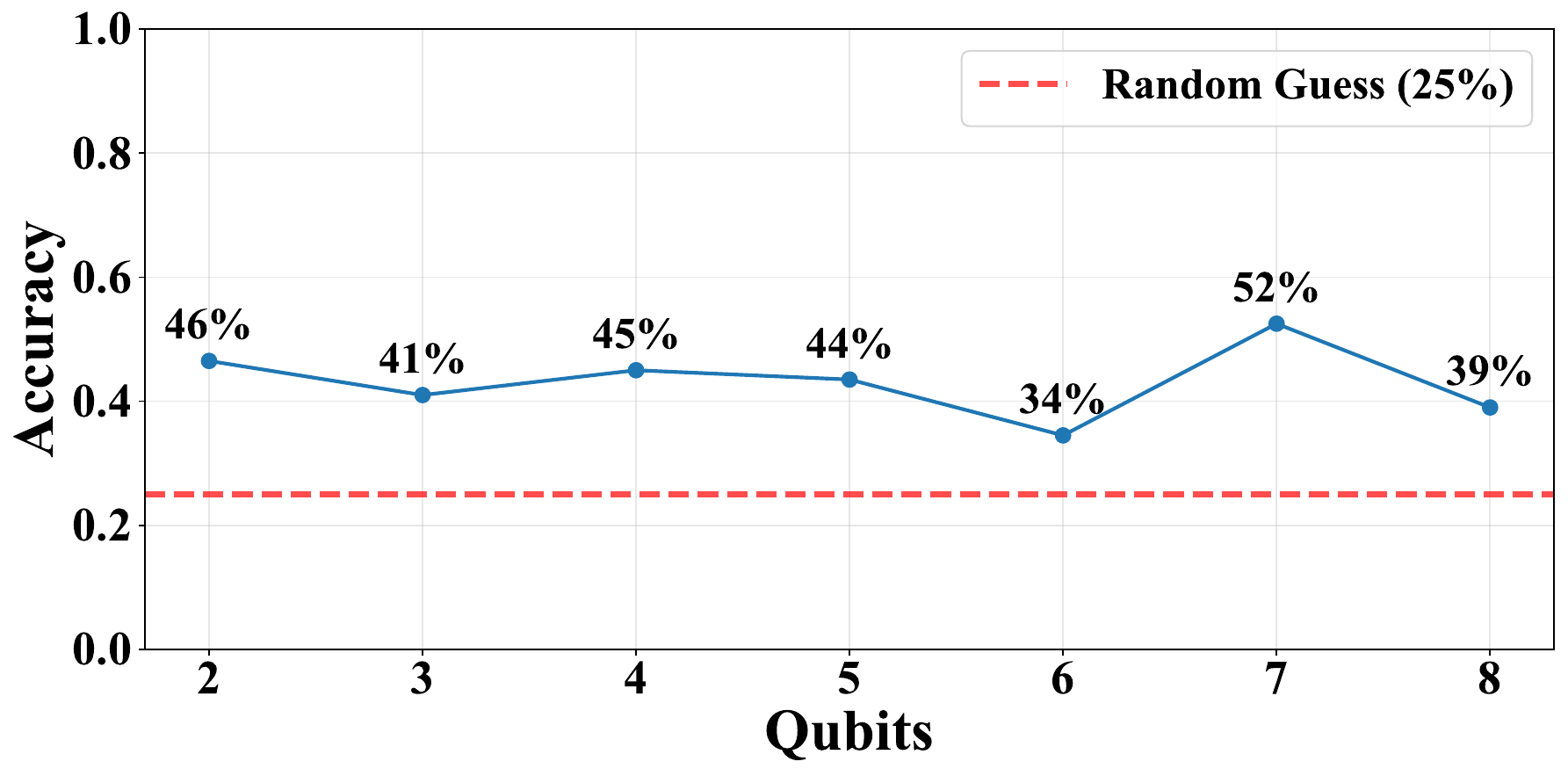}
    \caption{Large \texttt{qwen2.5} checkpoints versus the compact \texttt{qwen3:0.6b}. The 32B and 72B models remain above 0.53 accuracy even at 8 qubits, while the 0.6B baseline tops out near 0.52 on easy settings and falls to 0.34 on 8 qubits, emphasizing the benefit of pairing scale with verification-aware data.}
    \label{fig:model-row9}
\end{figure*}

\begin{figure*}[t]
    \centering
    \includegraphics[width=0.32\textwidth]{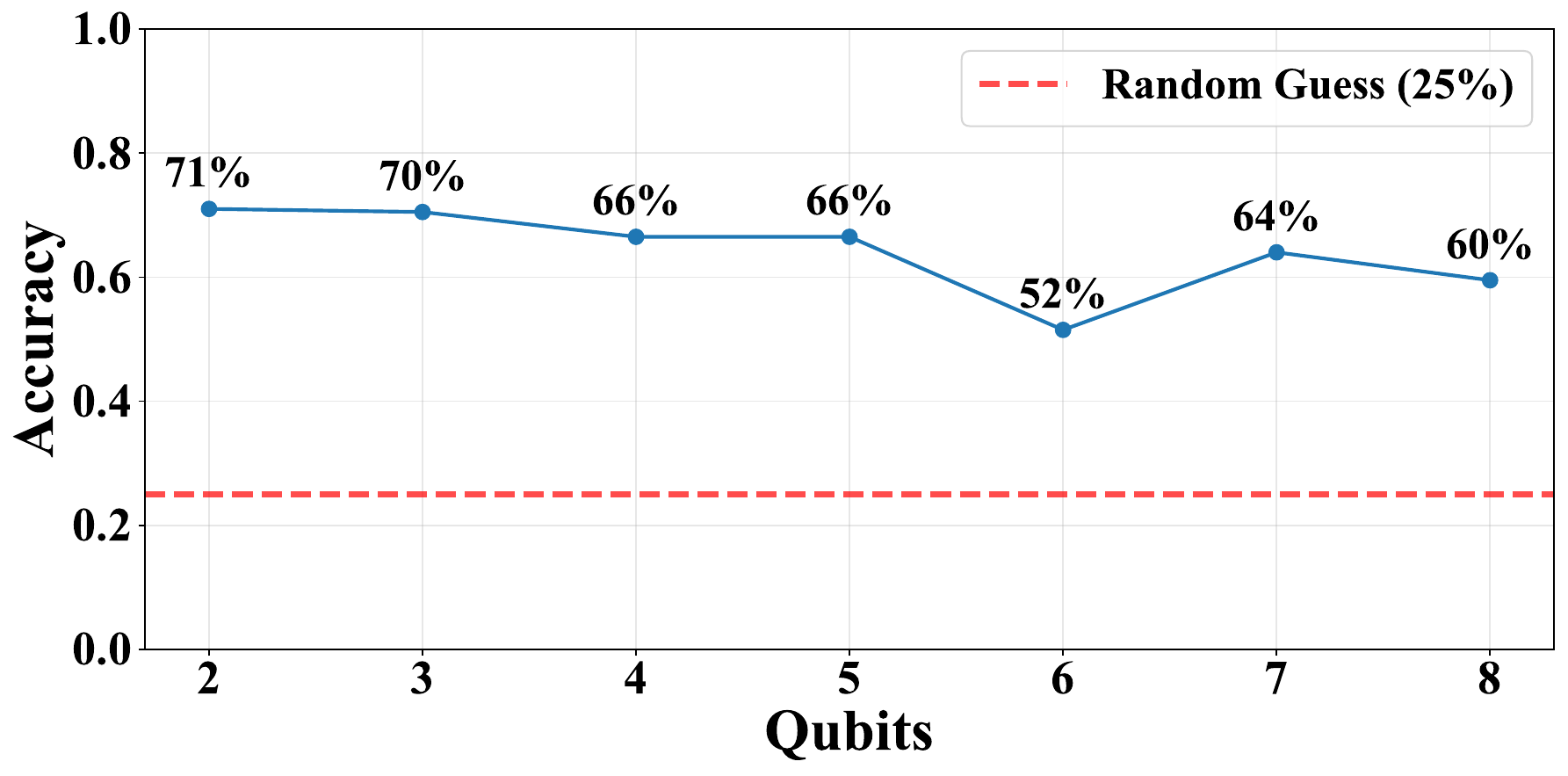}
    \includegraphics[width=0.32\textwidth]{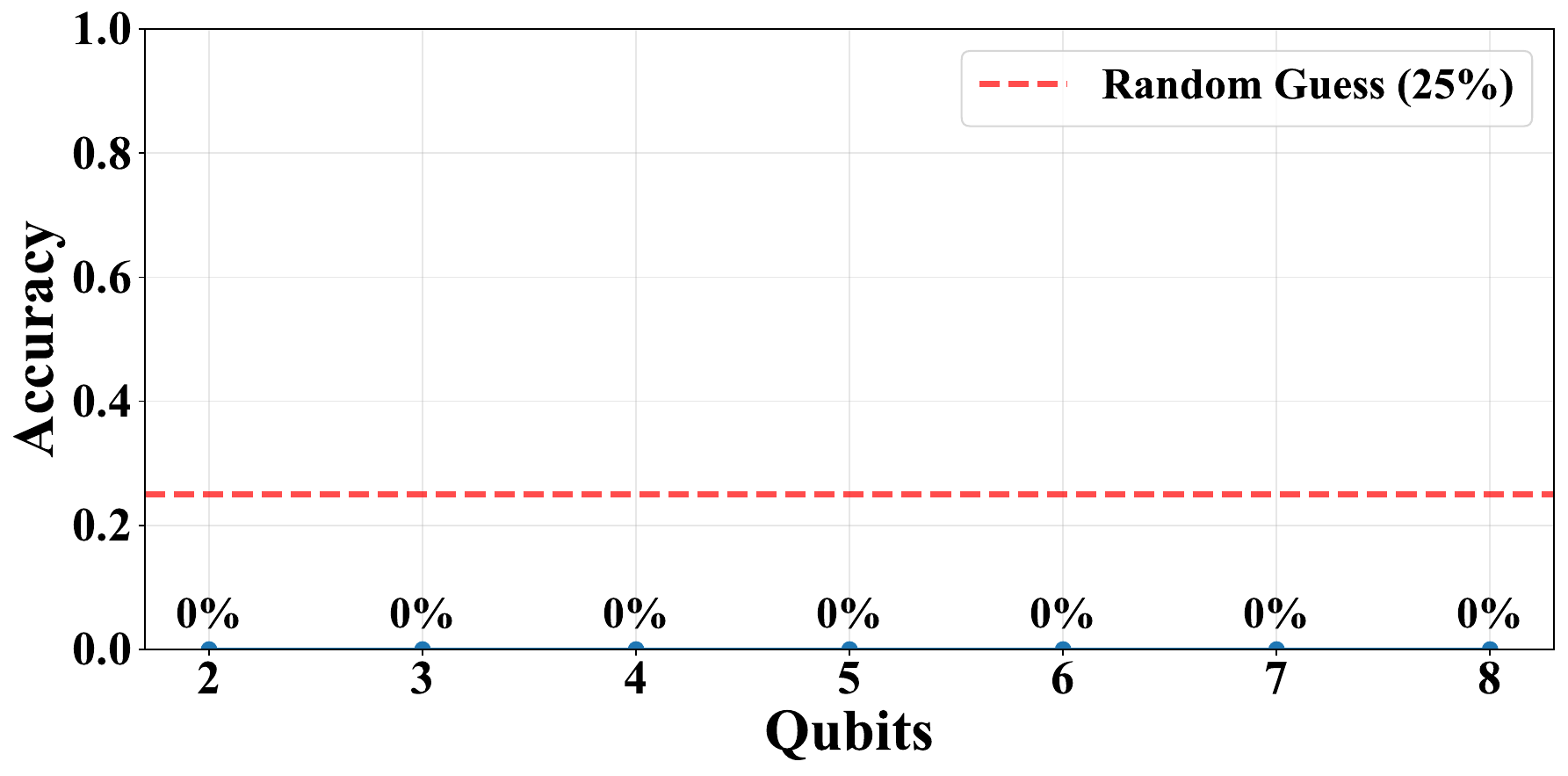}
    \includegraphics[width=0.32\textwidth]{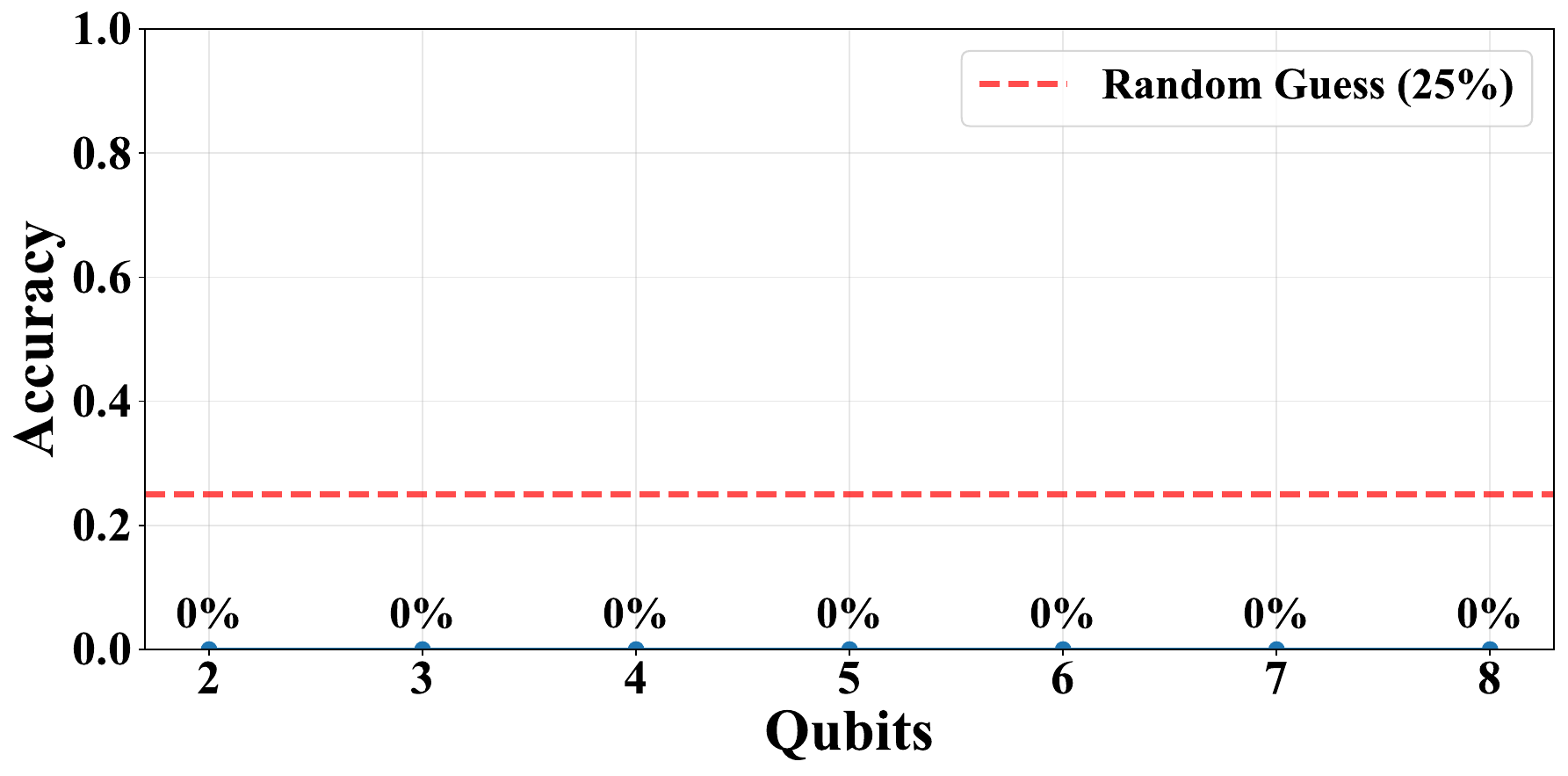}
    \caption{\texttt{qwen3} family. The 1.7B reasoning model maintains 0.52--0.71 accuracy across qubit counts, yet the larger 4B and 8B general-purpose releases answer zero questions correctly, highlighting how differing alignment objectives can erase MAJ/UMA skill despite added capacity.}
    \label{fig:model-row10}
\end{figure*}

\begin{figure*}[t]
    \centering
    \includegraphics[width=0.32\textwidth]{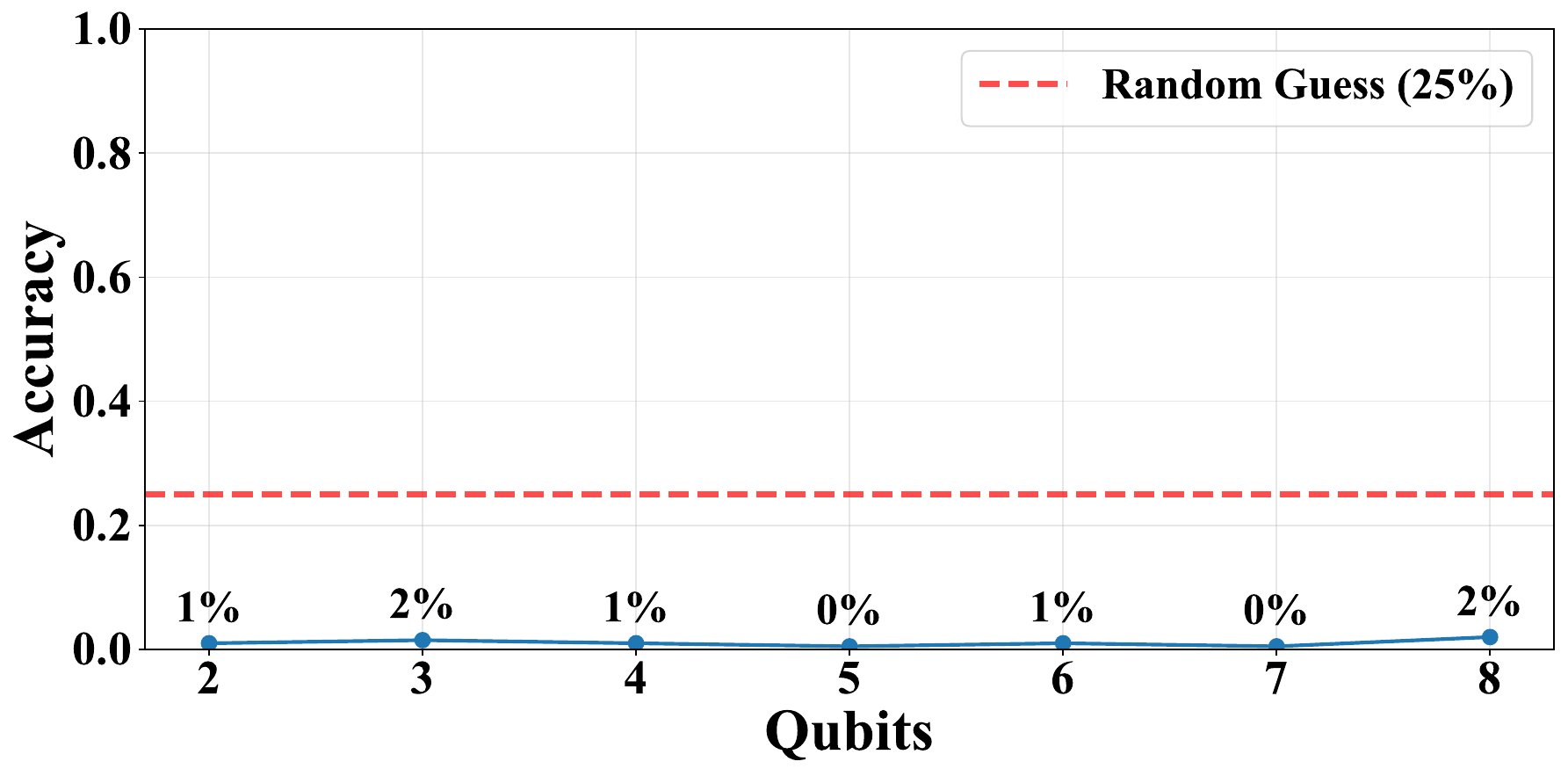}
    \includegraphics[width=0.32\textwidth]{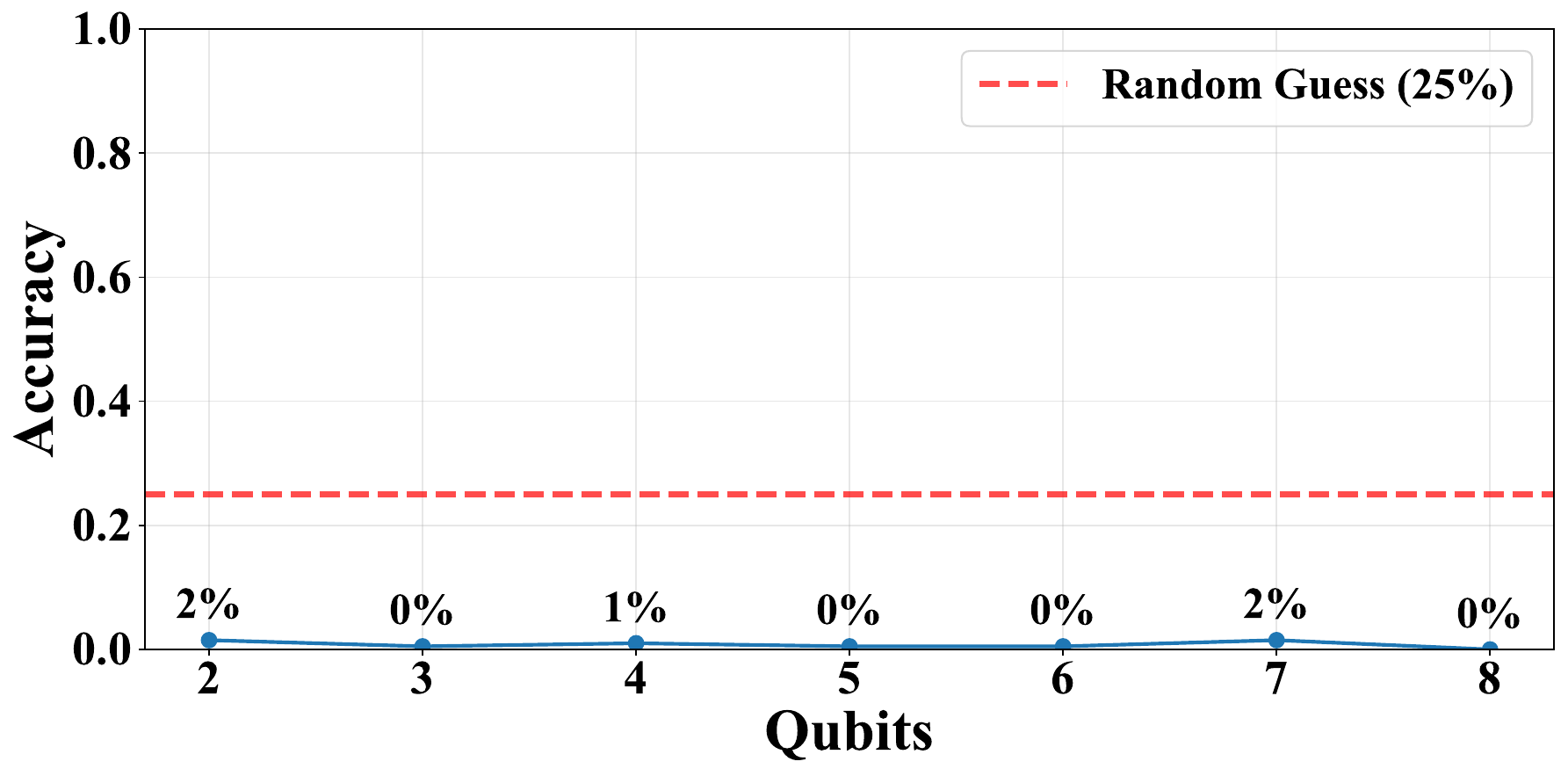}
    \includegraphics[width=0.32\textwidth]{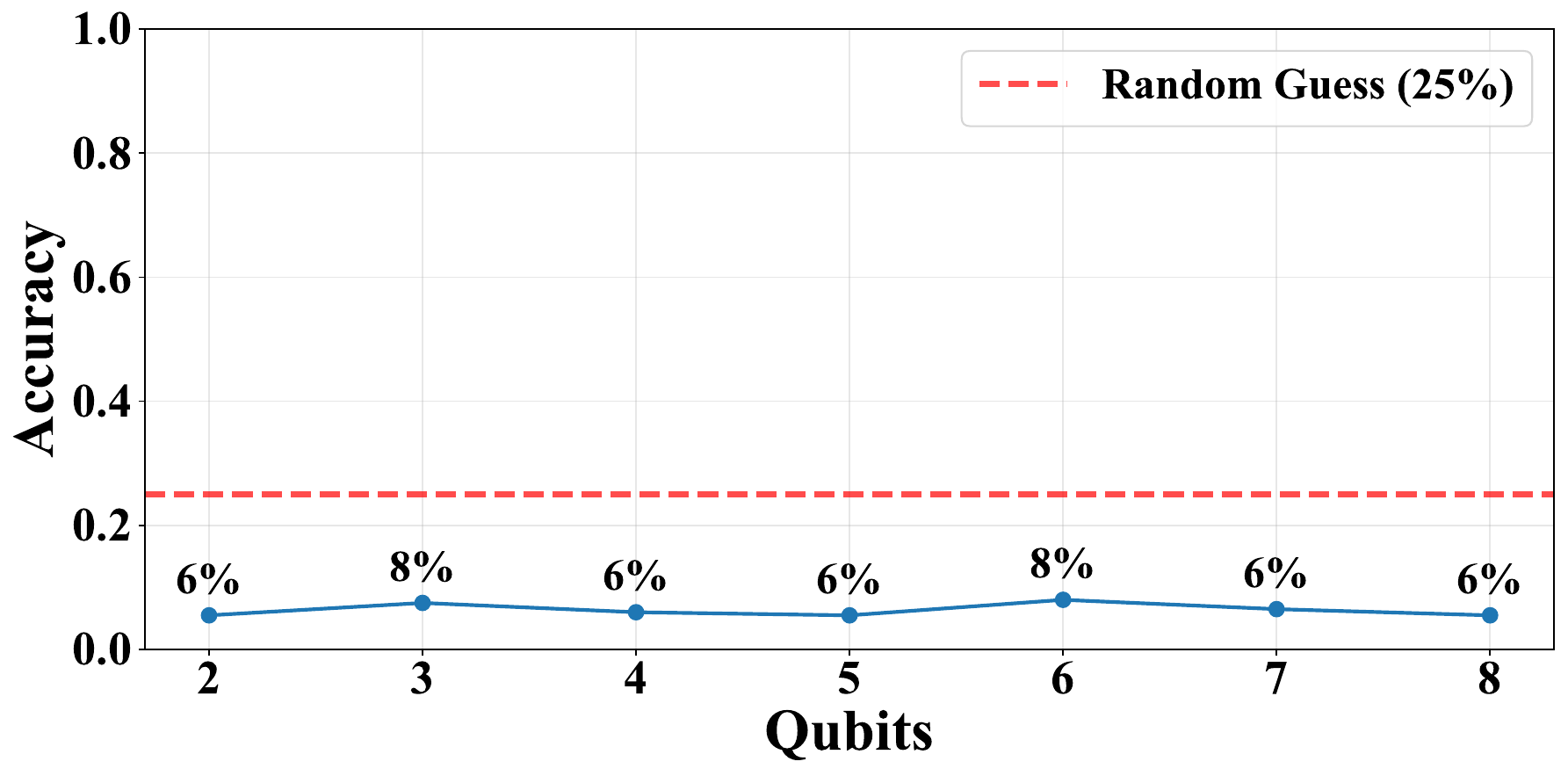}
    \caption{\texttt{deepseek-r1} models. Neither the 1.5B nor 7B variants exceeds 2\% accuracy, and even the newer 8B release plateaus at 6--8\%, confirming that exhaustive chain-of-thought generation alone does not translate into verifiable MAJ/UMA answers.}
    \label{fig:model-row11}
\end{figure*}

\textbf{Rationale for Multiple Visualization Dimensions.}
We provide complementary visualization types that each serve a distinct and irreplaceable analytical purpose. No single visualization can capture all aspects of model behavior; together they form a comprehensive evaluation framework:
\begin{itemize}
    \item \textbf{Heatmap} (Figure~\ref{app:heatmap}): Provides a \textit{global bird's-eye view} of all 34 models across all seven bit-widths in a single figure. The color gradient enables immediate pattern recognition---dark regions cluster around verification-aware models while lighter bands expose models stuck near random chance. This visualization is essential for identifying systematic trends (e.g., whether certain model families consistently outperform others) that would require extensive cross-referencing across dozens of separate plots.

    \item \textbf{Bit-width-grouped bar charts} (Figures~\ref{fig:bit-row1}--\ref{fig:bit-row4}): Enable \textit{horizontal cross-model comparison} at each fixed circuit complexity level. By sorting all 34 models by accuracy at a specific bit-width, readers can directly answer questions such as ``Which models perform best on 8-qubit circuits?'' or ``How many models exceed the random baseline at 5 qubits?'' The heatmap cannot answer these ranking questions as precisely, since color perception is less accurate than bar length comparison.

    \item \textbf{Model-grouped line charts} (Figures~\ref{fig:model-row1}--\ref{fig:model-row11}): Enable \textit{longitudinal scaling analysis} by tracking how each individual model's performance evolves as circuit complexity increases. These plots reveal critical insights invisible in aggregate views: some models maintain stable accuracy across bit-widths (indicating robust verification knowledge), while others exhibit sharp degradation beyond certain thresholds (suggesting brittle pattern matching). Such trajectory information is essential for predicting model behavior on larger, unseen circuits.

    \item \textbf{Radar chart} (Figure~\ref{fig:radar} in main text): Provides a \textit{compact aggregate summary} by averaging each model's accuracy across all bit-widths. While detailed breakdowns are informative, practitioners often need a single metric for quick model selection. The radar format visually separates verification-aware models (outer ring) from conventional copilots (clustered inside the 25\% baseline circle), enabling rapid identification of top performers.
\end{itemize}

In summary, these visualizations operate at different granularities: the heatmap shows everything at once but sacrifices precision; bar charts enable precise ranking but only at one bit-width per figure; line charts reveal temporal dynamics but only for one model per subplot; and the radar chart sacrifices detail for immediate comparative insight. Each visualization answers questions that the others cannot, making them collectively necessary rather than redundant. 


\end{document}